\documentclass{pasa}%

\title[Origin of evolution of transition discs]{The origin and evolution of transition discs: successes, problems and open questions}
\author[Owen, J. E.]{James E. Owen\thanks{E-mail: jowen@ias.edu}\thanks{Hubble Fellow}\\
\affil{Institute for Advanced Study, Einstein Drive, Princeton NJ, 08540, U.S.A.}}%
\jid{PASA}
\jyear{\the\year}

\usepackage[authoryear]{natbib}
\usepackage{color}
\bibpunct{(}{)}{;}{a}{}{,}
\newcommand{\bc}{}

\begin{document}%
\begin{abstract}
Transition discs are protoplanetary discs that show evidence for large holes or {\bc wide} gaps {(\bc with widths comparable to their radii)} in their dust component. These discs could be giving us clues about the disc destruction mechanism or hints about the location and time-scales for the formation of planets. However, at the moment there remain key gaps in our theoretical understanding. The vast majority of transition discs are accreting onto their central stars, indicating that - at least close to the star - dust has been depleted from the gas by a very large amount. In this review, we discuss evidence for two distinct populations of transition discs: mm-faint - those with low mm-fluxes, small holes ($\lesssim 10$~AU) and low accretion  rates ($\sim 10^{-10}-10^{-9}$~M$_\odot$~yr$^{-1}$) and mm-bright - discs with large mm-fluxes, large holes ($\gtrsim 20$~AU) and high accretion rates $\sim 10^{-8}$~M$_\odot$~yr$^{-1}$. MM-faint transition discs are consistent with what would naively be expected from a disc undergoing dispersal; however, mm-bright discs are not, and are likely to be rare and long-lived objects. We discuss the two commonly proposed mechanisms for creating transition discs: photoevaporation and planet-disc interactions, with a particular emphasis on how they would evolve in these models, comparing these predictions to the observed population. More theoretical work on explaining the lack of {\bc optically thick}, non-accreting transition discs is required in both the photoevaporation and planetary hypothesis, before we can start to use transition discs to constrain models of planet formation. Finally, we suggest that the few discs with primordial looking SEDs, but serendipitously imaged, showing large cavities in the mm (e.g. MWC758 \& WSB 60) may represent a hidden population of associated objects. Characterising and understanding how these objects fit into the overall paradigm may allow us to unravel the mystery of transition discs.       
\end{abstract}
\begin{keywords}
accretion, accretion disks --- protoplanetary disks --- planet-disk interactions --- planets and satellites: formation --- stars: pre-main sequence
\end{keywords}
\maketitle%
\section{INTRODUCTION}
\label{sec:intro}
Protoplanetary discs are a natural outcome of angular momentum conservation in star formation and are ubiquitous around young, forming stars. They represent the environment in which planets form, grow and migrate and any understanding of planet formation is intimately coupled to understanding the protoplanetary disc. 

The material that makes up the protoplanetary disc is sourced from the interstellar medium and as such is {\bc assumed to be} composed of 1~\% dust and 99~\% gas at birth, although, as we will discuss throughout this review the dust-to-gas ratio in the disc can evolve strongly in space and time during its lifetime. While the dust component is a minor constituent by mass it overwhelmingly dominates the opacity, to such an extent that the disc remains optically thick out to long wavelengths ($\sim$mm) and low disc masses. 

\begin{figure}
\centering
\includegraphics[width=\columnwidth]{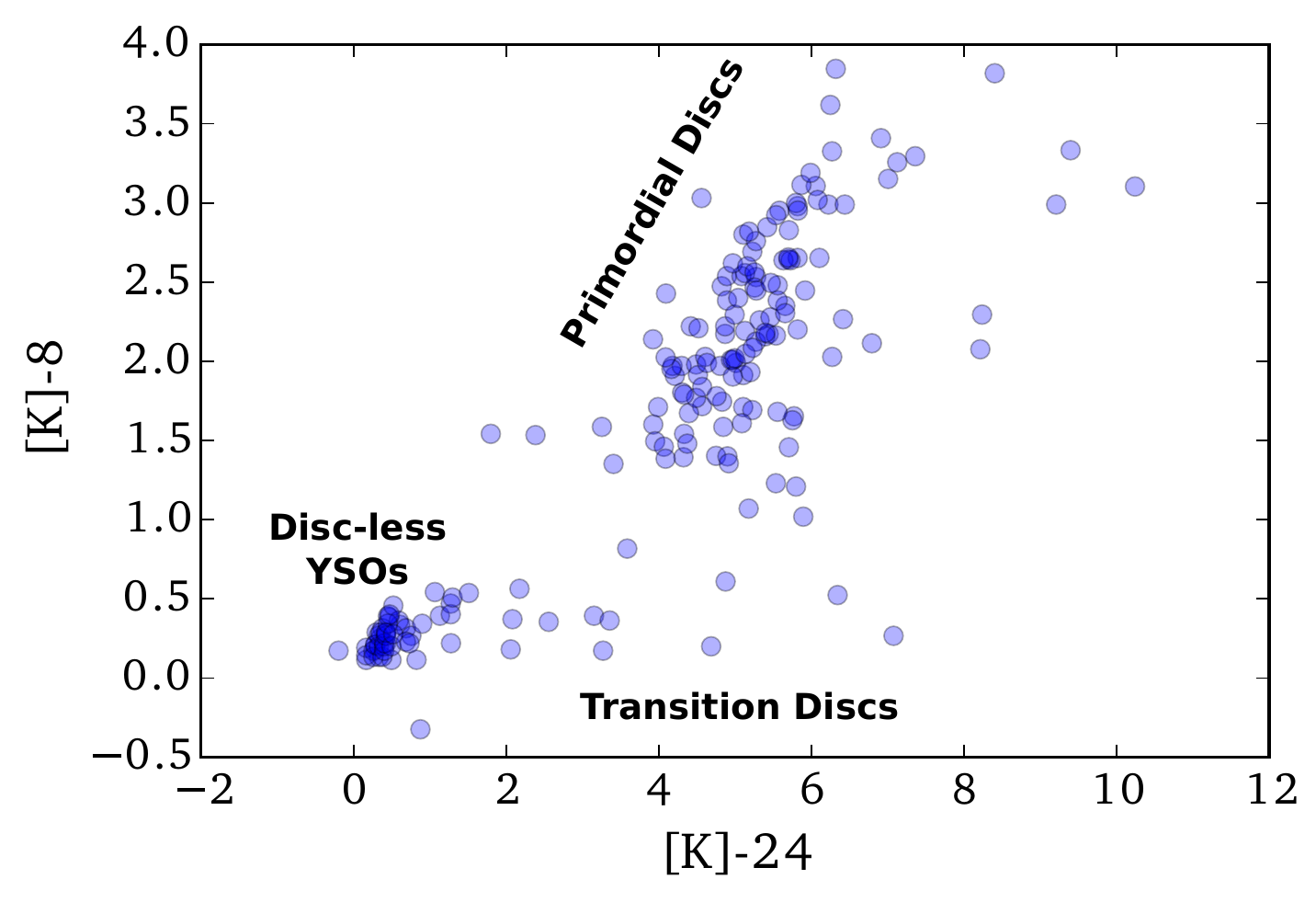}
\caption{{\it Spitzer} colour-colour diagram for young stellar objects with spectral-type range K4.5-M2.5 in nearby the star forming regions: Taurus, Tr37, Lupus, and Upper Sco. The different groupings described in the text are labelled. Data from \citep{Luhman2010,Koepferl2013}.}\label{fig:colour-colour}
\end{figure}

The fraction of young stars that show a near infra-red (NIR) excess \citep{Haisch2001,Hernandez2007,Mamajek2009} as a function of time indicates that at an age of $\lesssim1$~Myr most young stars harbour a disc; by $\sim 3$~Myr about 50~\% of stars have lost their discs and by $\sim10$~Myr almost every young star has lost its disc. The distribution of young stars observed  in the NIR colour versus MIR colour plane indicates two distinct separate groupings (see Figure~\ref{fig:colour-colour}): one consistent with stellar photospheres and hence disc-less stars, and another consistent with optically thick protoplanetary discs \citep[e.g.][]{Luhman2010,Koepferl2013} showing strong excess emission above the photosphere at NIR and MIR wavelengths, which we will refer to as ``primordial'' discs. Since most young stars sit in either of these two groupings, with very few objects not appearing as disc-less photospheres or optically thick discs \citep[e.g.][]{Strom1989,Skrutskie1990} then the time-scale it takes to disperse a primordial disc must be considerably shorter than the disc lifetime \citep[e.g.][]{Kenyon1995}.

However, there are a few objects that do not belong to the primordial or disc-less groupings.  These objects show no strong NIR excess, but MIR excesses consistent with the primordial discs \citep[e.g.][]{Strom1989,Skrutskie1990,Calvet2005,Espaillat2014}, indicating a lack of disc emission from small radii, but normal optically thick disc emission at large radii \citep{DAlessio2005,Ercolano2011,Koepferl2013}. Since these discs occupy an intermediate region in the colour-colour plane between the primordial discs and disc-less stars it has been hypothesised that such discs are caught in the act of dispersing, or ``transitioning'' between primordial discs and disc-less stars and as such have been described as transition discs \citep[e.g.][]{Kenyon1995}. We caution that while there is no evidence any transition discs are {\it actually} in the process of dispersing their disc, and as we will discuss later many may not be, the lack of intermediate objects in the colour-colour plane requires that disc dispersal must be rapid \citep[e.g.][]{Ercolano2011,Koepferl2013}. 

If we assume that when {\it every} disc disperses it follows a locus in the colour-colour plane such that it presents as a transition disc while it is dispersing and {\it every} observed transition disc is a one that is undergoing dispersal, then we can draw two important conclusions about disc dispersal. Firstly, as transition discs show a lack of NIR excess but show a MIR excess, one can infer that disc dispersal proceeds from the inside out. Secondly by comparing the number of transition discs to primordial discs we can measure the time-scale over which the dispersal takes place. Detailed studies have shown that for the range of spectral types F to M,  the transition disc fraction is $\sim10\%$ \citep{Kenyon1995,Luhman2010,Ercolano2011,Koepferl2013}. Coupled with a median disc lifetime of 3~Myr \citep[e.g.][]{Haisch2001,Hernandez2007,Mamajek2009}, one finds a disc dispersal time-scale of $\sim 0.3$~Myr. 

The observations of transition discs have lead to the development of many models to describe them, both within the framework of disc dispersal and outside it, such as: photoevaporation \citep[e.g.][]{Clarke2001}, planet-disc interactions \citep[e.g.][]{Calvet2005}, dust grain growth \citep[e.g.][]{Dullemond2005}, photophoresis \citep[e.g.][]{Krauss2007}, MRI-driven winds \citep[e.g.][]{Suzuki2009}, binary star interactions \citep[e.g.][]{Marsh1992} and MRI-driven dust depleted flows \citep[e.g.][]{Chiang2007}, with photoevaporation and planet-disc interactions considered the two most likely candidates. With the exception of dust grain growth all models try and explain the deficit in NIR excess emission by removing dust particles and in some cases gas from the inner regions, leaving behind a large hole or cavity in the dust disc with a radius ($R_{\rm hole}$), with observed hole radii spanning $\sim$1-100~AU. Understanding the origin of transition discs could help us understand not only disc dispersal but also planet-disc interactions and dust evolution. These are key issues for planet formation models and for which good observational constraints are lacking.

While many observational studies and associated reviews \citep[e.g.][]{Espaillat2014} have focused on identifying the origin of individual transition discs, relatively few have focused on describing how models predict how transition discs evolve and compare them to the population of transition discs as a whole. The aim of this review is to discuss this aspect and the implications the population of transition discs implies for the models and their {\it evolution}. We will explicitly ignore recently resolved non-axisymmetric features as this will be discussed in the accompanying review by \citet{Casassus2015c}. Thus, we will consider the two most commonly invoked models for creating a transition disc: photoevaporation and planet-disc interactions. Both are models for which theories for an evolutionary path can be sketched out. We will discuss the current model outlines, successes, failings, areas for further work and discuss what future observations will help unravel the transition disc mystery.

\section{INSIGHTS FROM OBSERVATIONS}
\label{sec:obs}
Transition discs were first identified through IR photometry \citep[e.g.][]{Strom1989} and to this date the majority of transition discs are studied purely through their spectral energy distributions (SEDs) \citep[e.g.][]{Calvet2005,Kim2009,Cieza2008,Merin2010,Espaillat2010,Kim2013}. The SED only gives information on the dust component and as such we must make inferences about the gas component.
\subsection{What constituents a transition disc?}
There is currently no consensus on what the definition of a transition disc is, nor is there likely to be in the future. The original idea of a disc that appears to be optically thin at NIR wavelengths and optically thick at MIR wavelengths appears to hold true to this day. Some authors \citep[e.g.][]{Currie2009,Sicilia2011} have argued that discs which show weaker excesses compared to the median SEDs taken from regions like Taurus are evidence of transition discs, which were sometime described as ``homologously depleted'' or ``anaemic'' discs and were possibly associated with a long disc dispersal phase. However, \citet{Ercolano2011} and \citet{Ercolano2011b} pointed out such structures arise from considering optically thick discs around late-type stars where the NIR excess is significantly weaker than that around a ``standard'' T-Tauri star due to the cooler stellar temperatures and smaller disc emitting area at NIR wavelengths \citep{Ercolano2011}. \citet{Koepferl2013} performed a detailed set of radiative transfer calculations as a function of spectral type and compared the observed data to the model disc structures. The only discs that were not optically thick at all wavelengths {\bc either} showed a lack of NIR emission while returning to optically thick values in the MIR, consistent with the original picture of a transition disc, {\bc or showed optically thin emission around Weak-lined T Tauri stars \citep[e.g.][]{Wahhaj2010,Cieza2013} that were consistent with the expectations of young debris discs.} 

Furthermore, \citet{Espaillat2007}, \citet{Kim2009} and \citet{Espaillat2010} showed that a fraction of those discs with colours and SEDs consistent with those similar to transition discs also contained a small amount of NIR emission consistent with either optically thin dust (as in GM Aur - \citealt{Espaillat2010}) or an optically thick but radially thin wall close to the dust-destruction front (as in UX Tau - \citealt{Espaillat2007}. \citet{Espaillat2010} labelled these discs as ``Pre-transition discs'', such that the small amount of inner dust could represent the final stages of its disappearance. To date there is no evidence that these discs are precursors to transition discs or that there is an evolutionary path that connects these discs to transition discs that show only photospheric emission out to the MIR. 

\begin{figure}
\centering
\includegraphics[width=\columnwidth]{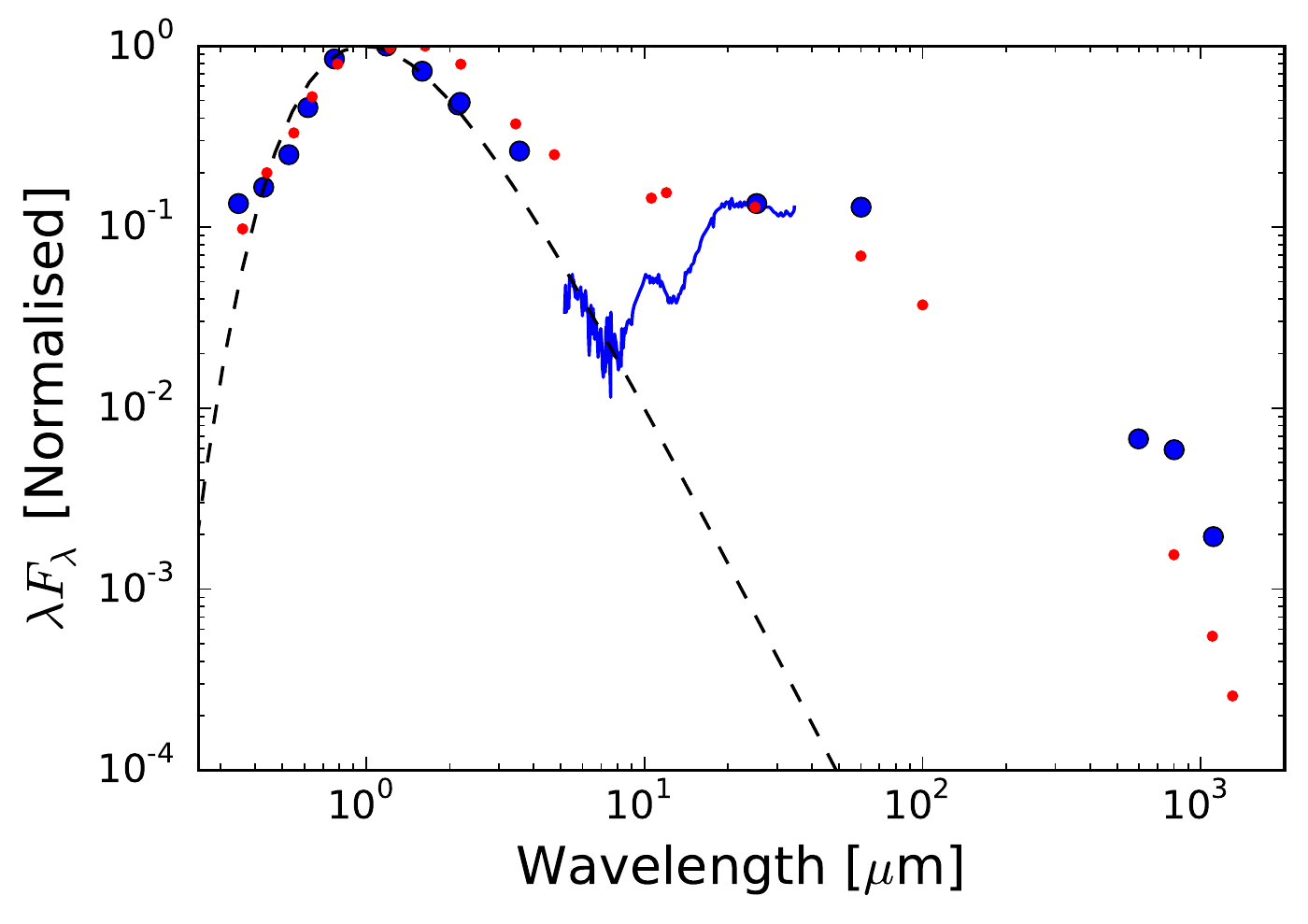}
\caption{The SED for the median primodial disc in the Taurus region \citep{DAlessio1999} is shown as the red points. The SED for the transition disc DM Tau is shown as blue filled circles \citep{Calvet2005,Furlan2009} and its IRS spectrum \citep{Calvet2005} as the blue solid line. A blackbody spectrum with $T_*=3720$~K is also shown as the dashed line representing the stellar photosphere for DM Tau \citep{Calvet2005}.}\label{fig:SED_compare}
\end{figure}

While there is no agreed definition of a transition disc, we choose to describe them qualitatively as discs which show evidence for a large depletion in opacity in the NIR, such that there is a large region of the disc close to the star that is optically thin to its re-radiated thermal emission. An example SED of a transition disc (DM Tau) compared to the Taurus median primordial SED \citep{DAlessio1999} is shown in Figure~\ref{fig:SED_compare} to demonstrate the transition disc signature. The transition disc DM Tau shows photospheric levels of emission out to $\sim 10$~$\mu$m, whereas the primordial SED has a strong IR excess above the photosphere. The fact that the transition disc SED shows a stronger Far IR (FIR) excess compared to the primordial SED is also evidence of a lack of NIR opacity, such that stellar irradiation that would normally be absorbed in the inner disc of a primordial disc is instead absorbed and re-irradiated by the outer disc \citep{Ercolano2015b}.

We refrain from labelling any disc that shows structure in imaging (e.g. HL Tau - \citealt{HLTAU2015}) but emission characteristics that do not match the general properties of transition discs described above. While such structures maybe related to the same mechanisms that give rise to transition disc structures this is not guaranteed, and lumping these objects in with transition discs distracts from their interesting and distinctly different characteristics. 

\subsection{Properties of the dust component from photometry}
Until recently almost all the information we have about transition discs came from continuum observations, either through photometry or imaging. Since the dust component of a protoplanetary disc dominates the opacity by many orders of magnitude, the information we can learn from observations primarily tells us about the structure and distribution of dust in the disc. 

What was realised early on was that the lack of NIR excess translated into a lack of dust at emitting temperatures from $\sim 1000$~K to a few hundred K. For sun-like stars these temperatures translate into radii of $\lesssim 10$~AU. Therefore, the primary feature of a transition disc is a {\it large} hole in the dust distribution in the inner region of the disc. We caution that a small gap with width $\ll R$ would be indistinguishable from an optically thick disc with no gap using pure photometric measurements. Therefore, whatever removes the dust particles from the inner disc must do so {\it globally} over a large portion of the disc, rather than locally. For typical values in primordial protoplanetary discs the dust surface density in the inner regions must be suppressed by a factor of order $10^{-4}$ in order to explain the observed NIR emission \citep[e.g.][]{Zhu2012}. 

A large number of transition discs show strong 10~$\mu$m silicate emission features \citep{Kim2009,Espaillat2010,Kim2013}. Such an emission feature can only come from small $<$1~$\mu$m sized dust particles inside the main dust hole, indicating that there is a small amount of optically thin dust in the hole, with a total mass content $<$ lunar mass \citep{Espaillat2010}. 

Furthermore, modelling of the transition between the optically thin hole and optically thick outer disc suggests it is sharp and does not happen gradually. Such a sharp transition generally rules out some evolutionary processes that may give rise to a transition disc signature such as grain growth \citep{Dullemond2005} which predicts a smooth variation from optically thin to thick \citep{Birnstiel2012}. 

\subsection{Properties of the gas component}

Compared to the relatively detailed information we can learn about the structure of the dust in the disc, the direct evidence we have for gas at certain locations in the disc is limited. For all discs we can look for evidence for gas very close to the star by determining whether the disc is actively accreting or not. Perhaps surprisingly the majority of transition discs are actively accreting \citep[e.g.][]{Kim2009,Kim2013,Espaillat2014,Manara2014}. While the average rates are slightly depressed from typical T Tauri rates \citep{Najita2007}, most discs are not accreting at very low rates with many having accretion rates of $\sim 10^{-8}$~M$_\odot$~yr$^{-1}$, and there is a large spread in accretion rates as discussed in section~\ref{sec:mdot_rhole}. 

Since the transport time-scales so close to the star are fast, and at minimum transition discs survive for $10^5$~years, then the gas that accretes onto the star cannot come from some small ($\lesssim 0.1$~AU) residual disc that resides close to the central star (such that any NIR excess is consistent with the observed values) and must be resupplied from some reservoir of at least ($M_{\rm res}\approx {\dot{M}_*/t_{\rm life}}$) of $\sim 0.1-1.0$ Jupiter masses.

Either this reservoir is already heavily depleted or the process that takes the material from large radius to small radius removes a large fraction of the dust. Therefore, in order to explain accreting transition discs there must be a strong separation of gas and dust in the inner disc. In the case of a standard viscous disc the suppression of the dust-to-gas mass ratio needs to be $\sim10^{-4}$ from nominal values \citep{Zhu2012}. 

Additionally, there are several observations that probe the gas component in the inner disc through molecular $\rho-$vibrational lines \citep[e.g.][]{Pontoppidan2008,Pontoppidan2010}. This indicates that there is a gas disc on scales of a few ~AU. Since a normal gas-to-dust ratio in such a gas disc would be eminently observable as a NIR excess, the presence of this gas further out from the star indicates that dust is strongly depleted from the gas in the inner regions.  

\subsection{Insights from imaging the gas and dust} 

Perhaps the biggest advance in the field has been the advent of imaging, particularly through sub-mm/mm continuum imaging of the dust component \citep[e.g.][]{Hughes2007,Brown2008,Hughes2009,Brown2009,Isella2010,Andrews2011}. Such imaging studies were the first to confirm - for the cases that were observable - that the interpretation of a transition disc SED as a disc with a large hole in its dust component was indeed correct, in many cases the observed hole size determined from imaging being consistent with that inferred from SED modelling. We should add that the SED and imaging do probe slightly different dust populations, with imaging sensitive to the mm-sized grains and the photometry sensitive to smaller grains (where the bulk of the opacity is). 

The largest sample of imaged transition discs was performed with the {\it SMA} by \citet{Andrews2011}, who studied 12 discs at 880~$\mu$m, measuring hole radii in the range 15-70~AU. Perhaps the most surprising aspect of the \citet{Andrews2011} sample is that it suggested the transition disc fraction at high mm fluxes (of all class II discs) was high $\gtrsim 25~\%$ and seemingly not what would be expected for a young star that is transitioning from disc-bearing to disc-less. We will return to this result in Section~\ref{sec:two_pop} where we discuss it in the context of {\it all} transition discs. Not only did this early observational imaging campaign confirm that the discs contained dust holes, but it also suggested that the mm emission was confined to a ring around the hole, rather than extending to many hundreds of AU, as seen in several other discs without large holes. This is suggestive of trapping of mm-sized dust particles \citep[e.g.][]{Pinilla2012a} in a pressure bump.

As {\it ALMA} came on-line we entered an era where high resolution sub-mm imaging of discs became possible. {\bc Currently, {\it ALMA} has not conducted a significant high resolution study of primordial discs, instead efforts have primarily concentrated on transition discs.} Surprisingly, this revealed that these dust rings were in some cases highly non-axis-symmetric. The most extreme case is IRS 48 \citep{vanderMarel2013}, which shows a azimuthal asymmetry of order ~100, with almost all the mm-sized dust particles being found on one side of the disc and none on the other. Such strong asymmetry is not seen in all discs, but IRS 48 is by no means the exception with SAO 206458 \citep{Perez2014}  also showing a strong azimuthal asymmetry. IRS 48 is an interesting object as it is one of the very few to have a MIR image as well \citep{Geers2007}. The MIR image of IRS 48 shows evidence for a hole in the dust disc but no asymmetry, but the sub-mm image shows evidence for a hole in roughly the same place but strong asymmetry. Since the MIR and sub-mm image probe different size dust particles, then not only do transition discs show strong dust and gas separations, but they also show strong separations between dust populations of different sizes. 

Additionally, {\it ALMA} has also opened up the ability to image the gas component of the disc. This has only been done in a handful of cases \citep{vanderMarel2013,Perez2015,vanderMarel2015,vanderMarel2015c}. The main result from these early studies was that the gas often extends inside the peak of the mm dust emission, but there is also evidence that there is a gas density drop inside the dust hole \citep{vanderMarel2015,vanderMarel2015c}. It is not clear how this fits into the entire picture yet as gas has only been imaged in a few cases and the interpretation is not without problems as we will discuss in Section-\ref{sec:planet}.

Finally, imaging has also been performed in the NIR using the PDI technique which is only sensitive to polarised and therefore scattered light. Such observations were pioneered by the SEEDS program \citep{SEEDSproject}; they have imaged many of the objects also imaged by the Andrews et al. (2011) sub-mm survey. Since these images are only sensitive to scattered light they are by construction tracing small particles ($\lesssim 1~\mu$m) in the surface layers of the disc at several scale heights. These scattered light images show a diverse range of characteristics with several discs showing evidence for spiral structure \citep{Muto2012}, dips in the brightness \citep{Thalmann2010}, holes in the image \citep{Mayama2012} as well as cases which show no structure and the scattered light images indicate emission from well inside the sub-mm hole \citep{Dong2012}. The varied nature of the observations makes it difficult to infer some general properties; however, again there is clear evidence of separations between small and large dust particles, with the small dust particles appearing to extend closer to the central star than the mm size dust particles \citep{Dong2012}. However, either the small dust particles are just tracing a small underlying population that is optically thin everywhere, or the small dust particles must be removed before they make it too close to the central star such that they would give rise to a strong NIR excess.   

\subsection{Two populations of transition discs}
\label{sec:two_pop}
The result of \citet{Andrews2011} that the fraction of {\it all} class II discs that were transition discs at high mm fluxes is high is intriguing. With the transition disc fraction in the upper quartile of all class II discs mm flux distribution being $\gtrsim 25$\%, this result is completely opposite to what one would expect from a young star that is transitioning from disc-bearing to disc-less. This is because if one assumes that mm-flux is a good proxy for disc mass, and disc mass declines with time due to accretion, then those young stars that are undergoing the transition from disc-bearing to disc-less should appear with the lowest mm fluxes in the class II distribution. Therefore, if all transition discs represented a class of discs undergoing the same dispersal mechanism at the end of their lifetimes then one would expect a monotonically declining transition disc fraction with increasing mm flux. This is exactly opposite to what was seen only at high mm fluxes where the transition disc fraction actually increased \citet{Andrews2011} as one went to higher mm fluxes.

\begin{figure}
\centering
\includegraphics[width=\columnwidth]{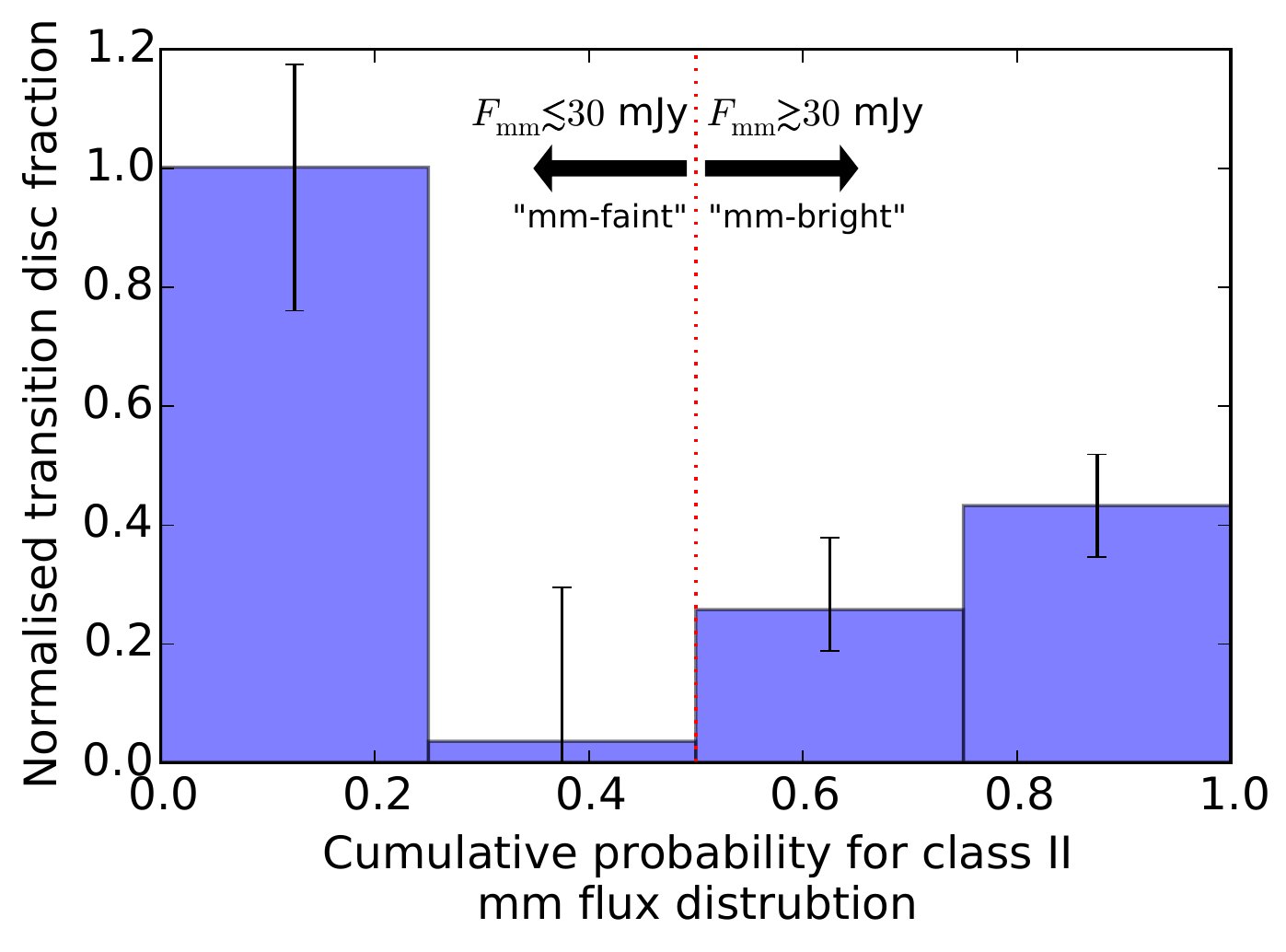}
\caption{The relative ratio of transition to primordial discs in each quartile of the primordial discs' mm distribution \citep{OC12}. The primordial discs mm distribution has a median 1.3mm flux of $\approx 30$~mJy at a distance of 140pc. }\label{fig:two_pop}
\end{figure}

However, as \citet{Andrews2011} readily point out, their sample is highly biased to those discs with the brightest mm fluxes and they did not comment on the transition disc fraction in the lower half of the class II mm disc distribution. \citet{OC12} studied the distribution of the transition disc fraction as a function of the entire mm flux distribution, to understand what happens to the transition disc fraction at low mm fluxes. The result of their study is shown in Figure~\ref{fig:two_pop}, where the normalised (to the first bin) transition disc fraction as a function of the class II mm flux distribution is shown.

\citet{OC12} recovered the rise in transition disc fraction found by \citet{Andrews2011} as the mm flux increased at high mm fluxes. However, they also found a rise in the transition disc fraction at low fluxes, with the majority of discs at the lowest mm fluxes being transition discs. The lack of transition discs at median class II mm fluxes is suggestive of at least two distinct populations of transition discs: one at the highest mm fluxes and another at the lowest mm fluxes. \citet{OC12} took the transition to occur at the median 1.3mm flux of approximately $30$~mJy assuming a distance of 140pc. 

Statistical tests of the properties of the hole sizes and mass accretion rates showed that these two populations had two distinctly different hole size and accretion rate distributions. The population at low mm fluxes have low accretion rates $\lesssim 10^{-9}$~M$_\odot$~yr$^{-1}$ and small hole sizes $\lesssim 20$~AU and the population at high mm fluxes have higher accretion rates $\sim10^{-8}$~M$_\odot$~yr$^{-1}$ and larger hole sizes $\gtrsim 20$~AU. 

\begin{figure}
\centering
\includegraphics[width=\columnwidth]{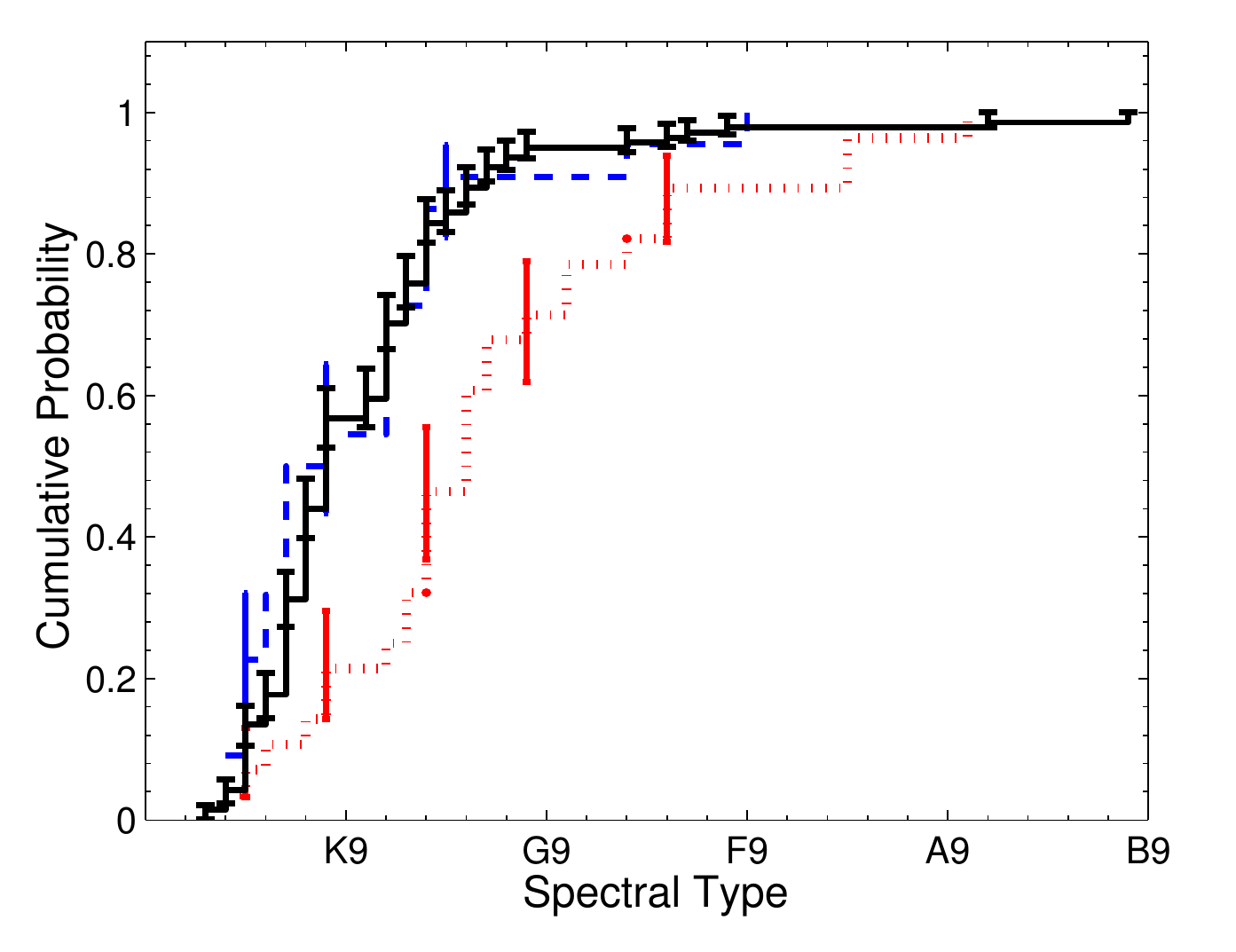}
\caption{The distribution of spectral types for the primordial discs (solid), mm-faint transition discs and mm-bright transition discs (dotted). Reproduction of Figure~3 from \citet{OC12}.}\label{fig:spec_compare}
\end{figure}

\citet{OC12} hypothesised that the populations at low mm-fluxes was actually a population of young stars in the process of dispersing their discs: i.e. actual {\it transition} discs, since they had all the evolutionary hallmarks of discs at the end of their lives: low disc masses and accretion rates. Furthermore, if all discs went through this process they argued that this low-mm flux population of transition discs should be uniformly drawn in spectral type for the entire population of the discs. Figure~\ref{fig:spec_compare} shows the spectral type distributions of low-mm flux transition discs (dashed), all discs (solid) and high-mm flux transitions discs (dotted). While it is clear the spectral type distribution of low-mm transition discs is indistinguishable from the entire population of discs, the high-mm flux population is clearly biased towards earlier spectral-types to a high level of significance (4.8$\sigma$) and therefore more massive stars.

In summary, there is strong evidence for two distinct populations of transition discs; here we label transition discs with a low mm flux: {\it mm-faint} and a high mm-flux: {\it mm-bright}. MM-faint transition discs consists of  a population at low mm fluxes, uniformly drawn from the underlying population of all young stars with small holes sizes and low accretion rates. This type of transition disc is consistent with a disc that is in the process of dispersing, with a rapid $\sim 10^5$~year dispersal time. 
MM-bright transition discs are a population at high mm fluxes, more commonly found around earlier type stars; the population has large hole sizes and accretion rates comparable to standard T Tauri stars. This type of transition disc is inconsistent with the idea that it is dispersing the disc. This type of disc does not have to satisfy any time-scale arguments inferred from the colour-colour plane, therefore, it is likely that this type of transition disc is {\it rare} and consequently long lived, with lifetimes of $\gtrsim 10^6$~years. \citet{OC12} argued that there are comparable numbers of each type of transition disc, with a slight preference for mm-faint transition discs. 

\subsection{The $\dot{M}_*-R_{\rm hole}$ plane}
\label{sec:mdot_rhole}
\begin{figure}
\centering
\includegraphics[width=\columnwidth]{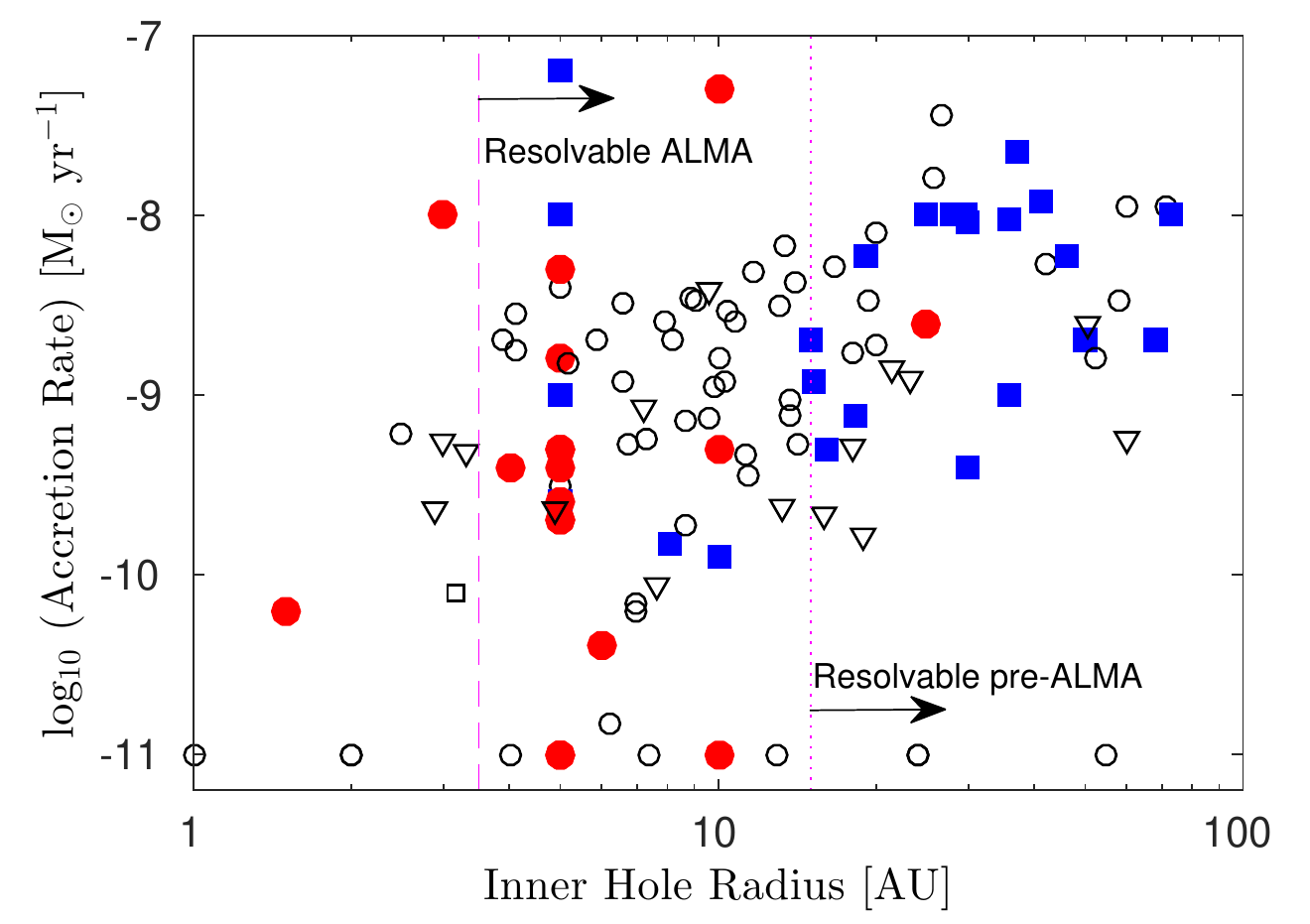}
\caption{The mass accretion rate as a function of inner hole size for transition discs. Blue squares are mm-bright transition discs, whereas red circles are mm-faint transition discs based on their mm flux. Open symbols do not have measured mm fluxes necessary to classify them and triangles represent upper limits in the accretion rate. Sample compiled from \citet{Calvet2002,Calvet2005,Najita2007,Espaillat2007,Espaillat2008,Cieza2008,Ercolano2009,Hughes2009,Kim2009,
Hughes2010,Najita2010,Merin2010,Cieza2010,Espaillat2010,Andrews2011,Andrews2012,Kim2013}. The vertical dotted line shows the hole radius of the smallest inner hole size directly imaged before {\it ALMA}. The vertical dashed line shows the hole radius resolvable with {\it ALMA}, assuming the same resolution as the HL-Tau observation \citep{HLTAU2015} at 140pc.}\label{fig:mdot_rhole}
\end{figure}

Perhaps one of the most interesting range of parameters to study transition discs in is the accretion rate hole size plane. As a transition disc evolves it should trace out a path in this plane and it is a useful diagnostic tool for testing evolutionary models of transition discs \citep[e.g.][]{Owen2011,Owen2012,OC12,Rosotti2013,Bae2013}. Figure~\ref{fig:mdot_rhole} shows a compilation of observed transition discs (taken from \citealt{Calvet2002,Calvet2005,Najita2007,Espaillat2007,Espaillat2008,Cieza2008,Ercolano2009,Hughes2009,Kim2009,
Hughes2010,Najita2010,Merin2010,Cieza2010,Espaillat2010,Andrews2011,Andrews2012,Kim2013}). MM-faint transition discs are the red circles, mm-bright transition discs are blue squares and those discs with no information on the mm flux are labelled with open symbols. While there is clearly a large scatter in accretion rate and hole size properties, this plane allows us to tease out important features. MM-faint transition discs clearly occupy a different region of parameter space to mm-bright transition discs, with mm-faint transition discs having lower accretion rates and smaller holes sizes. Furthermore, the majority of transition discs are accreting, with those transition discs that show no detectable accretion having small $\lesssim 10$~AU hole sizes. In the pre-{\it ALMA} era the smallest resolved hole  with sub-mm imaging was $\sim 15$~AU. This region does not correspond to the majority of transition discs and is completely dominated by mm-bright transition discs. However, assuming the same resolution as the HL Tau image \citep{HLTAU2015} can be achieved with future {\it ALMA} campaigns, then this resolution limit drops to approximately 3.5~AU at the distance of Taurus (140pc). This resolution means the vast majority of holes could be resolved with {\it ALMA}.     

There is no clear evidence for strong correlations; there is perhaps a tentative correlation that suggests hole size increases with accretion rate in mm-bright transition discs. However, this may only be driven by stellar mass rather than a specific evolutionary path. 

As we shall see the strongest constraint on any model is the significant lack of {\bc optically thick} transition discs with large $\gtrsim 20$~AU holes and no detectable accretion. There could plausibly be a population of ``relic'' discs with low-masses, large holes, {\bc but which are still optically thick} and no accretion rate that have been missed by current observations \citep{Owen2011}; {\bc however, the recent study by \citep{Hardy2015} suggest this possibility is rapidly disappearing and more extended {\it ALMA} surveys should easily rule out this possibility}. {\bc There are a small fraction ($\lesssim 20$\%) of weak-line T-Tauri stars (WTTs) that do show a infra-rad excess in {\it Spitzer} \citep{Padgett2006,Cieza2007,Wahhaj2010} and {\it Herchel} \citep{Cieza2013}. However, these objects general show a lack of emission from gas \citep{Pascucci2006,Ingleby2009} and dust at sub-mm wavelengths \citep{Duvert2000,Andrews2005,Andrews2007,Matthews2012}. There is a well known population of objects that show infra-red excesses until very late times (Gyrs), which are characterised by optically thin dust emission and the lack of gas (or very high dust-to-gas ratios), referred to as debris discs. In optically thin discs with a very low gas content, the small dust particles are removed by stellar radiation pressure or Ponting-Robinson drag on short time-scales. As such these debris discs are thought to represent discs in which a population of large km-sized bodies are colliding releasing small dust particles (and in some cases gas) that then give rise to optically thin infra-red excesses \citep[see][for a review]{Wyatt2008}. At late times there must be some transition between when primordial dust and gas from the protoplanetary disc is dispersed and secondary dust and gas from planetesimal collisions in debris discs is generated. Obviously, this could give rise to confusion into how to classify a WTTs with an infra-red excess: is it a transition disc at the end of its life or a young debris disc?

\citet{Cieza2013} argued that those WTTs stars that showed the strongest infra-red excesses, those discs perhaps consistent with an optically thick disc, were probably still accreting at a low-level or intermediately, and as such they were previously mis-identified as non-acrreting objects. However, those WTTs stars that showed infra-red excesses more consistent with optical thin emission similar to debris discs were unlikely to be accreting. Additionally, \citet{Hardy2015} studied a sample of WTTs with ALMA in both continuum and $^{12}$CO, detecting the continuum in 4/24 objects and CO in none, with inferred gas masses $\ll 0.1$~M$_J$, using the \citet{Williams2014} grid of models. Therefore, \citet{Hardy2015} concluded that the majority of WTTs with infra-red excesses were in-fact young debris discs rather than evolved transition discs. Taking as an upper limit all the \citet{Hardy2015} continuum detections to be evolved transition discs then this puts the fraction of stars with non-accreting transition discs with large holes close at around $\sim 3$\% considerably smaller than the fraction of young stars accreting optically thick transition discs. Therefore, any model should avoid the production of non-accreting, optically thick transition discs and either rapidly transition to a debris disc phase, or a gas-poor optically thin phase that is indistinguishable form a debris disc signature based on photometry alone \footnote{Since any secondary gas created by collisions or photo-dissociation of planetesimals will be volatile poor, this could plausibly be used as a observationally distinguishing characteristic between evolved transition discs and debris discs.}, although the exact fractions are still uncertain due to small number statistics.

}

\subsection{Uncertain Issues}

There are still several areas where we have limited information. MM-bright transition discs are the least well studied. In most cases all the information in hole size and dust distribution come from photometry and SED fitting. Furthermore, apart from the measured accretion rates there are no observations that probe whether the dust holes host a gas disc. Additionally, given the hole sizes are small and mm fluxes are low, no imaging campaign has confirmed the presence of the holes inferred from the SED or probed any non-axisymmteric structure. Since these discs are likely to actually be discs in transition, more observations are required to constrain the structures in both gas and dust of these discs.

MM-bright transition discs are comparatively well studied, especially those which were bright enough to be early {\it ALMA} targets. However, there are still a few outstanding issues, particularly with regard to the structure and kinematics of the gas component and the distribution of small dust particles in the outer disc. For example: are the bulk of the small dust particles removed at a location similar to the mm-sized dust particles or do they extend much further in as hinted at by SEEDs images of the surfaces of the discs? MIR imaging of the thermal dust component appears to be the way forward to answering this question \citep[e.g.][]{Geers2007}.

\section{THE PHOTOEVAPORATION SCENARIO}
\label{sec:photo}
Photoevaporation of accretion discs occurs whenever the surface layers of discs are heated to sufficient temperature such that they can drive a pressure driven, transonic wind that escapes the pull of gravity. Photoevaportive disc winds were first discussed in the context of Compton heated winds from discs surrounding black holes \citep{Begelman1983a,Begelman1983b} and later in the solar system \citep{Shu1993} and discs around young massive stars \citep{Hollenbach1994}. These early works considered the winds working alone to clear the disc, with clearing time-scales estimated simply by taking the disc mass and dividing it by the mass-loss rates, to find:
\begin{equation}
t_{\rm clear}=100~{\rm Myr} \left(\frac{M_{\rm d}}{0.1~{\rm M}_\odot}\right)\left(\frac{\dot{M}_w}{10^{-9}~{\rm M}_\odot~{\rm yr}^{-1}}\right)^{-1}.
\end{equation}
This approach yielded very long time-scales, so long that it was thought that photoevporation played no role in the evolution of protoplanetary discs around low-mass stars \citep{Hollenbach1994}. However, it was realised by \citet{Clarke2001} that while photoevaporation could remove a limited amount of mass from the disc, when combined with viscous evolution it could have a dramatic effect on the disc. \citet{Clarke2001} suggested that when combined with viscous evolution, photoevaporation could reproduce the two time-scale nature of disc dispersal that proceeds from inside out, as indicated by the observations discussed in Section~\ref{sec:obs}. 

The photoevaporative switch operates as follows: consider a disc with $\dot{M}_*\gg\dot{M}_w$ that is evolving due to viscous evolution and photoevaporation. In order to accrete onto the star any fluid parcel in the outer disc must lose angular momentum due viscous processes and radially move from the outer disc to the inner disc. Since $\dot{M}_*\gg\dot{M}_w$ there is an extremely small chance that the fluid parcel is removed from the disc by the photoevaporative wind, rather than being accreted onto the star, thus the disc behaves as a normal viscously evolving protoplanetary disc. However, as the disc viscously drains onto the central star the accretion rate through the disc drops; eventually, one will reach the point where the accretion rate in the disc at large radius equals the photoevaporation rate. Therefore, any fluid parcel that starts viscously drifting in will always be taken out of the disc by the photoevaporative wind rather than accreting onto the star. This means that the wind starves the star and inner disc of resupply eventually opening a gap in the disc. Such a schematic picture also tells us where the gap will first open in the disc, since the gap will happen at the last point a fluid parcel has a chance of being taken out in the wind, i.e. the radius in the disc where the wind is first launched\footnote{One has also to consider whether the decrection flow of the outer edge of the inner disc is weak enough that it cannot partially refill the gap; in all practical cases this condition has little consequence}.

This radius can be estimated from an energetic argument, by comparing the thermal energy of the gas to its binding energy in the star's potential, which yields a ``gravitational radius'' ($r_g$), following \citet{Hollenbach1994} as:
\begin{eqnarray}
r_g&=&\frac{G M_*}{c_s^2},\nonumber \\&=&8.9~{\rm AU}\left(\frac{M_*}{1~{\rm M}_\odot}\right)\left(\frac{c_s}{10~{\rm km~s}^{-1}}\right)^{-2}.
\end{eqnarray}
Numerical simulations \citep{Font2004,Owen2010} have shown that the wind can be launched sub-sonically from a slightly closer radius, and this has been called the ``critical radius'' by \citet{Alexander2006a,Alexander2006b} which has a position:
\begin{equation}
r_{\rm cr}\approx 2~{\rm AU}\left(\frac{M_*}{1~{\rm M}_\odot}\right)\left(\frac{c_s}{10~{\rm km~s}^{-1}}\right)^{-2}.
\end{equation} 
Therefore, the expectation is that for a solar-like star photoevaporation will open a gap in the disc around an AU. 

\begin{figure}
\centering
\includegraphics[width=\columnwidth]{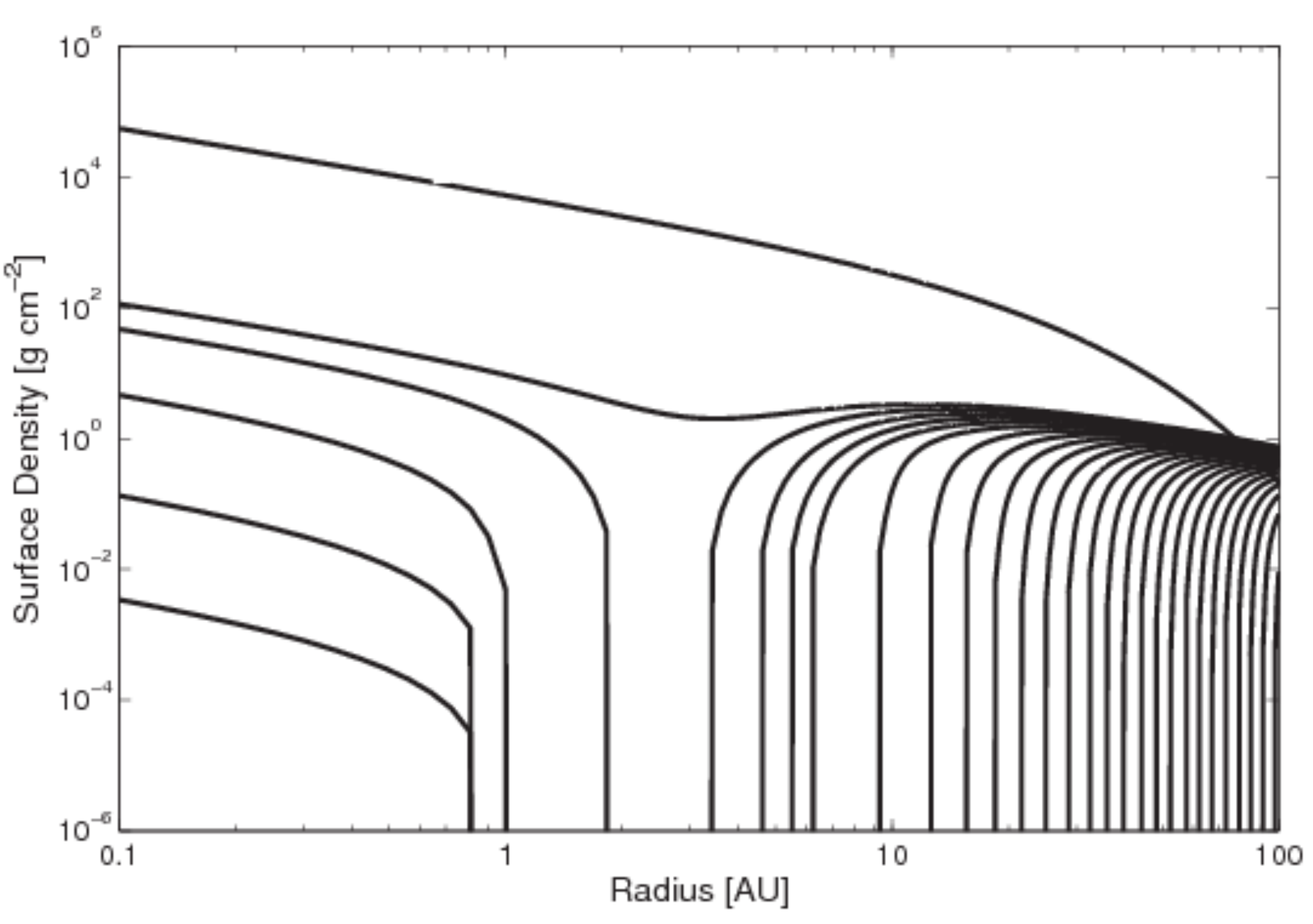}
\caption{Evolution of the surface density of a viscously evolving photoevaporating disc.  The  first  line  shows  the  initial surface  density  profile,  the next shows the profile at 75 per cent of the discs lifetime ($\sim$3.5 Myr) and the  remaining  lines  show  the  surface  density  at  1  per  cent  steps  in  disc lifetime. Reproduction of Figure~9 from \citet{Owen2011}.}\label{fig:owen2011}
\end{figure}

The viscous time-scale of a protoplanetary disc increases strongly with radius; i.e. for a constant $\alpha$, passively heated ($T\propto R^{-1/2}$ - \citealt{Kenyon1987}) protoplanetary disc the viscous time-scale grows linearly with radius \citep{Hartmann1998}. Therefore, this inner disc will evolve on a much shorter time-scale and drain onto the central star on a time-scale shorter than the current disc's lifetime. This leaves a {\it completely} clear gap in gas and dust, with a gas and dust rich disc outside the gap that is then eroded to large radius. The evolution of a disc undergoing photoevaporative disc dispersal is shown in Figure~\ref{fig:owen2011} taken from \citet{Owen2011} using the X-ray photoevaporation model \cite{Owen2011,OC12}. Here the gap opens after approximately 3.5~Myr at $\sim 2$~AU. The inner disc then drains onto the central star in a few $10^5$~years and subsequently the disc is eroded to larger radii in a further few $10^5$~years.   

While there are several flavours of the photoevaporation model: EUV \citep[e.g.][]{Hollenbach1994,Font2004,Alexander2006a,Alexander2006b}, X-ray \citep[e.g.][]{Alexander2004,Ercolano2008,Ercolano2009b,Owen2010,Owen2011,Owen2012} and FUV \citep[e.g.][]{Gorti2009a,Gorti2009b,Gorti2015} there is still debate within the photoevaporation community about the exact values of the mass-loss rates and the driving mechanisms \citep[see][for a recent reviews]{Alexander2014}. However, the production and evolution of transition discs created by photoevaporation is well established: once photoevaporation opens a gap the inner disc drains onto the central star in a few $10^5$~years and then the outer disc is then eroded to large radius. Although, in the case of large dead-zones this process can be more complicated \citep{Morishima2012,Bae2013}. 

\subsection{Transition discs created by photoevaporation}
Perhaps the most challenging aspect for the photoevaporation model is to explain accreting transition discs. If one assumes that dust follows the gas in Figure~\ref{fig:owen2011} then while the disc is accreting onto the star (even after the gap has opened) it would appear as a primordial disc through its SED, as a small gap from 1-5 AU would not present a sufficient suppression in IR excess to be identified as a transition disc. Such an argument was previously used to rule out photoevaporation as the origin of a transition disc, regardless of the accretion rate or hole size \citep[e.g.][]{Kim2009}. 

\begin{figure}
\centering
\includegraphics[width=\columnwidth]{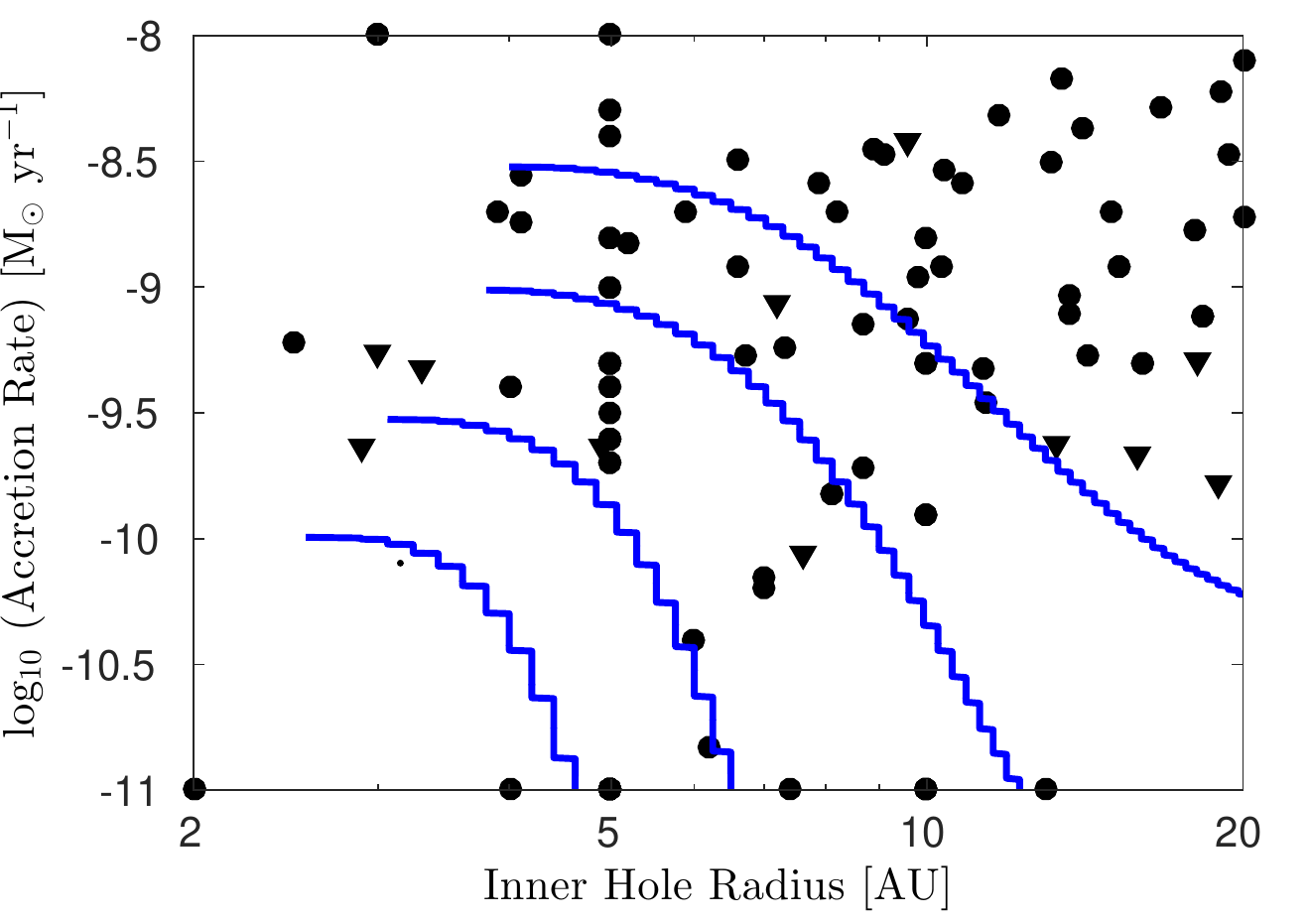}
\caption{Tracks of the accretion rate against hole size for accreting transition discs produced by the photoevaporation model, for accretion rates at gap opening of $3\times10^{-9}$, $10^{-9}$, $3\times10^{-10}$ \& $10^{-10}$~M$_\odot$~yr$^{-1}$ taken from the \citet{Owen2011} calculations. The points represent observed transition discs from Figure~\ref{fig:mdot_rhole}.}\label{fig:Owen2011_tracks}
\end{figure}

However, it is well known in protoplanetary discs that dust does not necessarily follow the gas (we shall see this argument throughout this article) as dust experiences dust drag \citep[e.g.][]{Weidenschilling1977,Armitage2010}. Once photoevaporation opens the gap in the gas disc preventing resupply of gas from the outer disc, it also prevents resupply of dust to the inner disc. Therefore, the dust rapidly spirals into the star under the action of gas drag on a time-scale of order $10^4$~years \citep{Alexander2007}. \citet{Owen2011}, argued that observationally this means that as soon as the disc opens a gap due to photoevaporation, observationally it would appear as an accreting transition disc with a dust hole from the star to a radius of $1-10$~AU (effectively instantaneously) after gap opening. {\it Therefore, there is no problem with photoevaporation producing an accreting transition disc signature.} However, as photoevaporation is a threshold process: i.e. the gap always opens around 1~AU when the accretion rate drops below the photoevaporation rate then by construction the accretion rate in the transition disc cannot be higher than the photoevaporation rate and the hole size is limited to the radius to which the disc can be eroded while the inner disc is draining onto the central star. Therefore, even with the most optimistic choices about photoevaporation, accreting transition discs are limited to $\dot{M}_*\lesssim 10^{-8}$~M$_\odot$~yr$^{-1}$ and $R_{\rm hole}\lesssim 20$~AU. In Figure~\ref{fig:Owen2011_tracks} we show tracks of the accreting transition discs created by the X-ray photoevporation model taken from \citet{Owen2011} for accretion rates at gap opening of $3\times10^{-9}$, $10^{-9}$, $3\times10^{-10}$ \& $10^{-10}$~M$_\odot$~yr$^{-1}$. These are representative of the range of possibilities photoevaporation can produce. Therefore, during this accreting transition disc stage the model predictions closely match the observed properties of mm-faint transition discs \citep[e.g.][]{OC12}. 

Once the inner disc has drained onto the central star the outer disc will be eroded to large radius, and will observationally appear as a non-accreting transition disc with a hole size anywhere between a few AU and 100s of AU. Since disc mass is dominated at large not small radii\footnote{For the canonical $\Sigma\propto R^{-1}$ surface density profiles \citep{Hartmann1998,Andrews2009}.}, all models where photoevaporation erodes the outer disc to large radii predict that this second phase, where the disc appears as a non-accreting transition disc, lasts a comparable or slightly longer amount of time than the accreting transition disc phase. The non-accreting transition disc population is dominated by discs with the largest hole sizes (i.e. the snapshots pile up at large radius in Figure~\ref{fig:owen2011}). {\bc If photoevaporation preferentially removes gas over dust (only very small dust particles can be entrained in the photoevaporative flow, \citealt{owen2011_dust}), then the dust-to-gas mass ratio will increase with time. However, globally this enhancement is limited: at gap opening the total mass in the disc is similar to the total gas mass removed in the wind. Thus, globally photoevaporation can only increase the dust-to-gas ratio by a factor of a few. Therefore, at large hole sizes the predicted remaining disc masses from the EUV \citep{Alexander2006b,Alexander2009}, FUV \citep{Gorti2009b,Gorti2015} and X-ray photoevaporation \citep{Owen2011} are above the sub-mm dust and gas mass limits for the majority of WTTs with infra-red excesses reported by \citet{Hardy2015} of $\lesssim 0.3$~M$_\oplus$ in dust and $\ll 0.1$ M$_J$ in gas. Even the lowest photoevaporation rates of $\sim10^{-10}$~M$_\odot$~yr$^{-1}$ from the X-ray photoevaporation model should give rise to transition discs that are optically thick out to very large hole radii $\gtrsim 100$~AU \citep{Owen2011}.}

Thus, the lack of observed {\bc optically thick} non-accreting transition discs with large hole radii, points to an issue with the photoevaporation model. \citet{Owen2012,Owen2013}, suggested that ``thermal-sweeping'', a dynamical instability that clears the disc once the inner disc has drained onto the central star, would remove this long-lived non-accreting transition disc phase, with no discs with holes larger than $30-40$~AU expected. However, \citet{Haworth16}, has cast doubts on the efficiency of X-ray driven ``thermal-sweeping'', and even the \citet{Owen2013} calculations over-predict the number of non-accreting transition discs as there are very few non-accreting transition discs with holes in the range 20-40~AU seen in Figure~\ref{fig:mdot_rhole}. As such, the lack of non-accreting transition discs from the photoevaporation model still remains an open problem and issue for the photoevaporation scenario.

\subsection{Summary}

\begin{figure*}
\centering
\includegraphics[width=\textwidth]{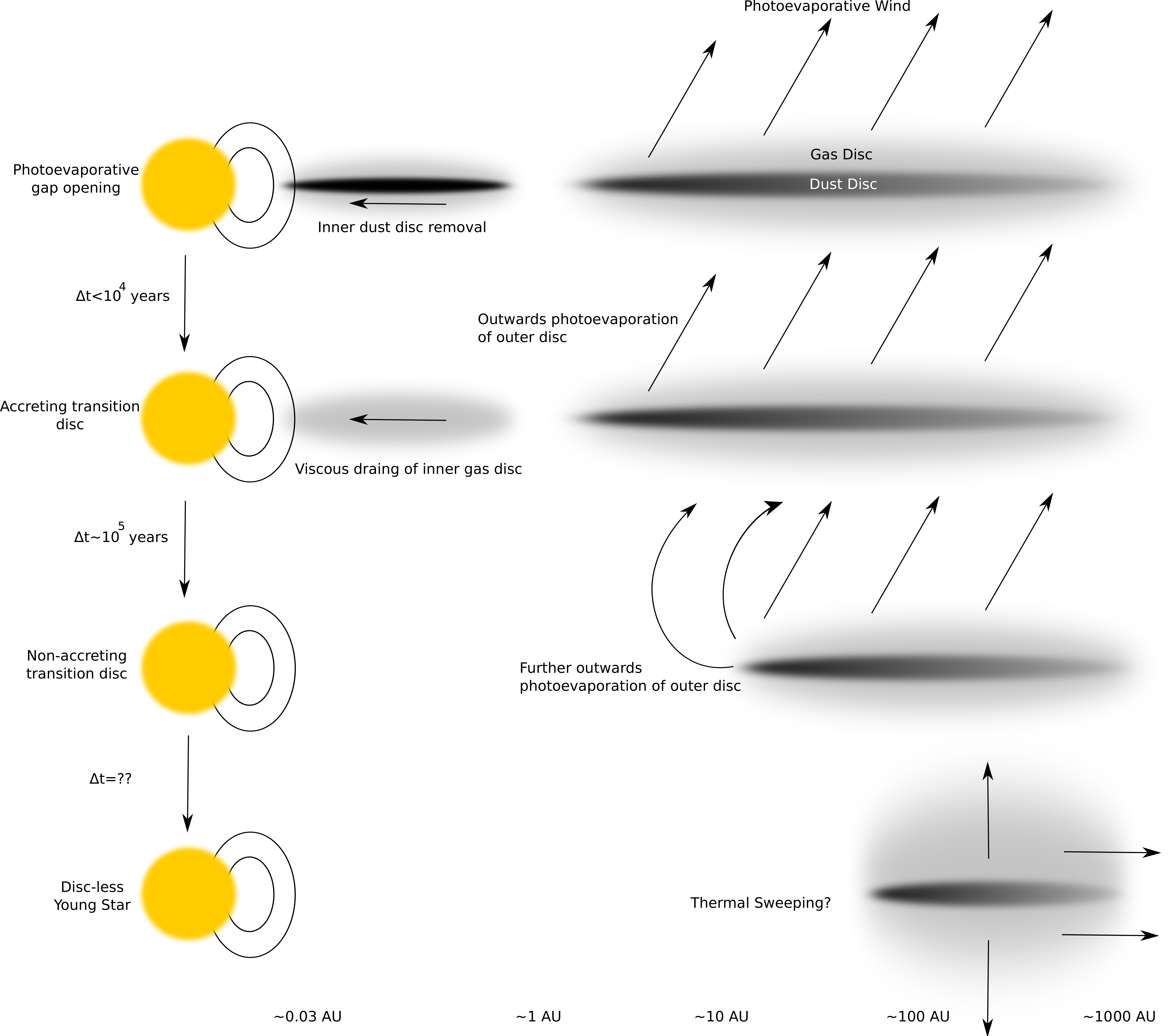}
\caption{Schematic picture of photoevaporation driven dispersal of a protoplanetary disc and the various transition disc stages (c.f. \citealt{Alexander2014}). After gap opening (stage I) the dust in the inner disc rapidly drains onto the star due to dust drag in $\lesssim 10^4$~years leaving an accreting transition disc (stage II). The inner dust-free gas disc viscouly drains onto the central star in a few $10^5$~years leaving a non-accreting transition disc (stage III) where photoevaporation erodes the disc to large radius where something like thermal sweeping finally destroys the disc leaving behind a disc-less yong star (stage IV). The lifetime of stage III is highly uncertain theoretically. The approximate radial scales are shown at the bottom for reference. Stage III is not observed. }\label{fig:photo_evolve_diagram}
\end{figure*}

Photoevaporation naturally creates a transition disc phase that occurs at the end of the disc's lifetime. This happens after the accretion rate through the disc drops below the photoevaporation rate opening a gap in the inner disc at a few AU of a solar mass star. The formation and evolution of a photoevaporation-created transition disc is shown schematically in Figure~\ref{fig:photo_evolve_diagram}. Once the gap is opened dust is quickly removed from the inner disc on a time-scale of $\lesssim 10^{4}$~years \citep{Alexander2007}. The inner - now dust free - disc continues to accrete onto the star with a lifetime of a few $10^5$~years appearing as an accreting transition disc signature \citep{Owen2011}, consistent with the properties of mm-faint transition discs\citep{OC12}. While the inner disc is draining onto the star, the outer disc is continually photoevaporating, such that the hole erodes to a radius $\lesssim 20$~AU. Once the inner disc has fully drained onto the central star photoevaporation becomes more efficient (due to the large exposed area at the inner edge of the gap) and the disc is photoevaporated to large hole sizes \citep{Alexander2006b,Owen2011}. The disc will now appear as a non-accreting transition disc with a large hole $\gtrsim~20$~AU until it is eventually destroyed possibly by ``thermal sweeping'' \citep{Owen2013}.

Photoevaporation can only form transition discs at the end of the discs' lifetime, with small hole sizes $\lesssim 10$~AU and low accretion rates $\lesssim 10^{-8}$~M$_\odot$~yr$^{-1}$. Therefore the photoevaporation model is consistent with the properties of mm-faint transition discs. Photoevaporation alone cannot explain any properties of mm-bright transition discs.

The photoevaporation model alone predicts a long non-accreting transition disc phase comparable with, if not longer than, the accreting transition disc phase \citep{Clarke2001,Alexander2006b,Owen2011}. This {\bc appears} inconsistent with the lack of observed {\it optically thick} non-accreting transition discs with large holes (see Figure~\ref{fig:mdot_rhole}) and remains a problem for the photoevaporation model, and in fact any disc dispersal mechanism. The number of {\bc possible optically thick}  non-accreting transition discs (or those with upper limits) is small compared to those with accretion $\lesssim 20$\% indicating that the non-accreting transition disc lifetime is short compared to the total transition disc lifetime and that outer disc dispersal is considerably shorter than inner disc dispersal (assuming it occurs from inside out). The ``thermal-sweeping'' instability \citep{Owen2013} may go a long way to resolving these issues; however, it is too early to say if it works efficiently enough to be consistent with the data. {\bc Alternatively, if grain-growth becomes efficient in these low-mass, dust rich, discs they could plausibly become optically thin, and thus be consistent with the population of WTTs with optically thin dust discs \citep[e.g.][]{Padgett2006,Cieza2007,Wahhaj2010}, which so far are hypothesised to be young debris discs \citep{Cieza2013,Hardy2015}, where the infra-red excesses arises from second generation dust.}

\section{THE PLANETARY HYPOTHESIS}
\label{sec:planet}
The origin of transition discs from an interaction between the disc and a forming planet is highly appealing. If we can understand the specific planet-disc interactions that lead to a transition disc signature, then we may be able to use the observations of transition discs to observationally probe planet formation. The idea that planet formation can also disperse the disc (by turning the disc material into planets rapidly - \citealt{Armitage1999}) has fallen out of favour. Not least because the time-scale for planet formation strongly increases with separation, appearing inconsistent with the rapid dispersal of the outer disc and furthermore because the disc lifetime - metallicity correlation \citep{Yasui2009} is inconsistent with planet formation and more likely explained by the X-ray photoevaporation model \citep{EC11}. Additionally, the giant planet fraction at large separations is low \citep[e.g.][]{Cumming2008,Brandt2014}, seemingly inconsistent with the idea all discs turned themselves into planets. However, this does not mean that planets are not associated with transition discs, just they are not the primary drivers of disc dispersal. Furthermore, the recent discovery of sub-stellar companions in several transition discs \citep{KI2012,Biller2012,Close2014,Sallum2015}, gives credence to the idea that transition discs maybe created by planet-disc interactions. 

It is well known that massive planets can carve gaps in protoplanetary discs, with the most basic idea being that the planet's Hill sphere is larger than the disc's scale height \citep{Lin1993}. This gives rise to the idea of the ``thermal gap opening mass'' ($M_{\rm th}$), obtained by equating the size of the hill sphere to the scale height of the disc. Which, for parameters approximating that required by transition discs, equates to:
\begin{eqnarray}
M_{\rm th}&=&\frac{c_s^3}{G\Omega},\nonumber\\
&\approx&0.2~{\rm M_J}\left(\frac{c_s}{0.4~{\rm km~s}^{-1}}\right)^3\left(\frac{M_*}{1~{\rm M}_\odot}\right)^{-1/2}\!\!\!\!\left(\frac{a}{20~{\rm AU}}\right)^{3/2}
\end{eqnarray}
while this criterion provides a lower limit in a very viscous disc, since viscosity can re-fill the gap \citep{Crida2006}, or an upper estimate in a disc with very low viscosity, where waves excited by very low mass planets can damp at a distance from a planet and open a gap \citep{Dong2011,Zhu2013,Duffell2013} it provides a convenient guide. Namely, a massive (gas giant planet) is probably able to open a gap in a protoplanetary disc. However for a single planet the gap width ($\Delta$) is contained to be within a few hill radii of the planet (the planets torque falls of with distance from the planet as $r^4$ - \citealt{Lin1979}) as such a single planet on a circular orbit can only open a gap of width \citep[e.g.][]{Fung2014}:
\begin{equation}
\Delta\approx2R_{\rm H}=5~{\rm AU}\left(\frac{a}{20~{\rm AU}}\right)\left(\frac{M_p}{1~{\rm M_J}}\right)^{1/3}\left(\frac{M_*}{1~{\rm M}_\odot}\right)^{-1/3}.
\end{equation}

Since at minimum, transition discs require dust material to be removed at least down to 0.1~AU from holes with sizes of 10s of AU, clearly the gas structure created by a single planet is not sufficient to give rise to a transition disc signature. 

\begin{figure}
\centering
\includegraphics[width=\columnwidth]{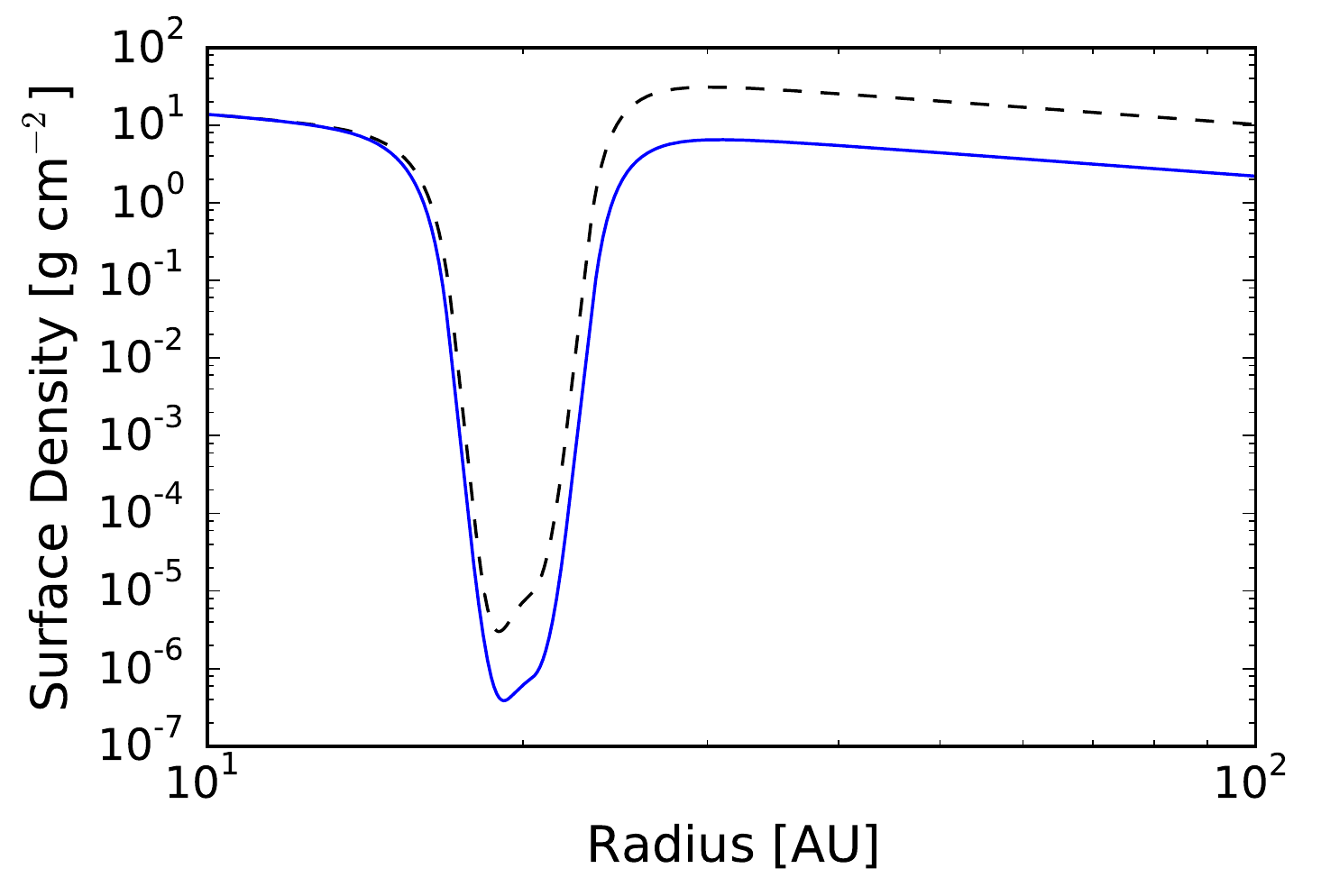}
\caption{Surface density profile of a disc containing an embedded 4.0~M$_J$ planet on a circular orbit at 20~AU. The dashed line shows the case with the planet accreting at 6 times the accretion rate onto the star, the solid line shows the case with the planet accreting at 1/2 the rate onto the star. In all cases the accretion rate onto the star is $\dot{M}_*=10^{-8}$~M$_\odot$~yr$^{-1}$.}\label{fig:gas_owen}
\end{figure}

Unlike photoevaporation which can only maintain accretion for a short while after creating the gap, this is not a problem for the planet model. While planets do create a gap, tidal torques from the planet also provide a mechanism for efficient angular momentum transport of gas across the gap in narrow high velocity streams. Thus, planets create a leaky gap. Many simulations of these planet disc interactions that account for accretion onto the planet and across the gap \citep[e.g.][]{Lubow2006}, suggest that the accretion rate into the inner disc (and hence onto the star) and the accretion rate onto the planet are comparable. This means for the accretion rates observed in transition discs with large holes ($10^{-9}-10^{-8}$~M$_\odot$~yr$^{-1}$), even though a planet may open a gap at sub-jovian masses, it will quickly accrete enough gas to become a Jupiter, or even a super-Jupiter. Furthermore, if the planet is accreting a significant amount of the gas that flows into the gap, then there should be a gas drop across the gap. The value of this drop should be consistent with the drop in accretion rate across the gap. If we assume the viscosity does not change in magnitude across a ({\it thin}) gap such that $\Sigma\propto \dot{M}$ then the change in gas surface density between the outer disc $\Sigma_{\rm out}$ and the surface density just inside the hole ($\Sigma_{\rm hole}$) is given by:
\begin{equation}
\frac{\Sigma_{\rm out}}{\Sigma_{\rm hole}}=1+\left(\frac{\dot{M}_p}{\dot{M}_{*}}\right).
\end{equation}    
Since the simulations suggest $\dot{M_p}/\dot{M_*}$ is of order unity then the change in gas surface density should be a factor of a few. The gas surface density profile as a function of radius is shown in Figure~\ref{fig:gas_owen} for an 4.0~M$_J$ planet at 20~AU accreting at 6$\dot{M}_*$ (dashed) and 1/2$\dot{M}_*$ (solid), showing the narrow gap produced by the planet, but a noticeable drop in gas surface density due to the planets accretion. 
Such a drop in gas surface density near the hole could be an observational signature of a planet hosting transition disc \citep{vanderMarel2015}. 

\subsection{Dust distribution}
While planets have the ability to create a long lasting accreting transition disc, we must still analyse how the structure induced in the disc by planets could results in a transition disc signature. Outside the orbit of the planet the gas surface density exhibits a maxima (see Figure~\ref{fig:gas_owen}), where viscous inflow of material is ({\it approximately}) balanced by the expulsion of material due to the planet's torque. The radial pressure gradient also balances the gas disc radially against gravity, therefore, the gas will not be orbiting with exactly the Keplerian velocity. Infact, the gas will orbit with a velocity:
\begin{equation}    
u_\phi=v_k\left[1+n\left(\frac{H}{R}\right)^2\right]^{1/2},
\end{equation}
where $n$ is the logarithmic radial pressure gradient (${\rm d}\!\log \!P/{\rm d}\!\log \!r$). Therefore, a positive pressure gradient ($n>0$) results in a super-Keplerian velocity and a negative pressure gradient ($n<0$) results in a sub-Keplerian velocity. The pressure maxima resulting from the planet's perturbation creates a structure where $n$ changes sign. Therefore, just outside the planet's orbit the gas is orbiting in a super-Keplerian manor. With increasing radius, one eventually goes past the pressure maxima the gas orbits in a sub-Keplerian fashion, where the pressure maxima typically occurs at twice the planet's separation \citep{Pinilla2012b}. Since dust particles have too high a density to feel any dynamical effect from the pressure gradient, they essentially orbit the star with a Keplerian velocity. As such any dust particles orbiting the star in a gas rich disc that has a super-Keplerian velocity will feel a tail wind, increasing their angular velocity and hence causing them to move outwards. For dust particles orbiting the star in a gas disc that is sub-Keplerian the opposite happens, they feel a head wind, which causes them to lose velocity and move inwards. Thus the effect of a gas disc is to cause dust particles to migrate towards pressure maxima. The time-scale at which this process happens is strongly linked to the drag force on the particles and unsurprisingly to the particle size.

\begin{figure}
\centering
\includegraphics[width=\columnwidth]{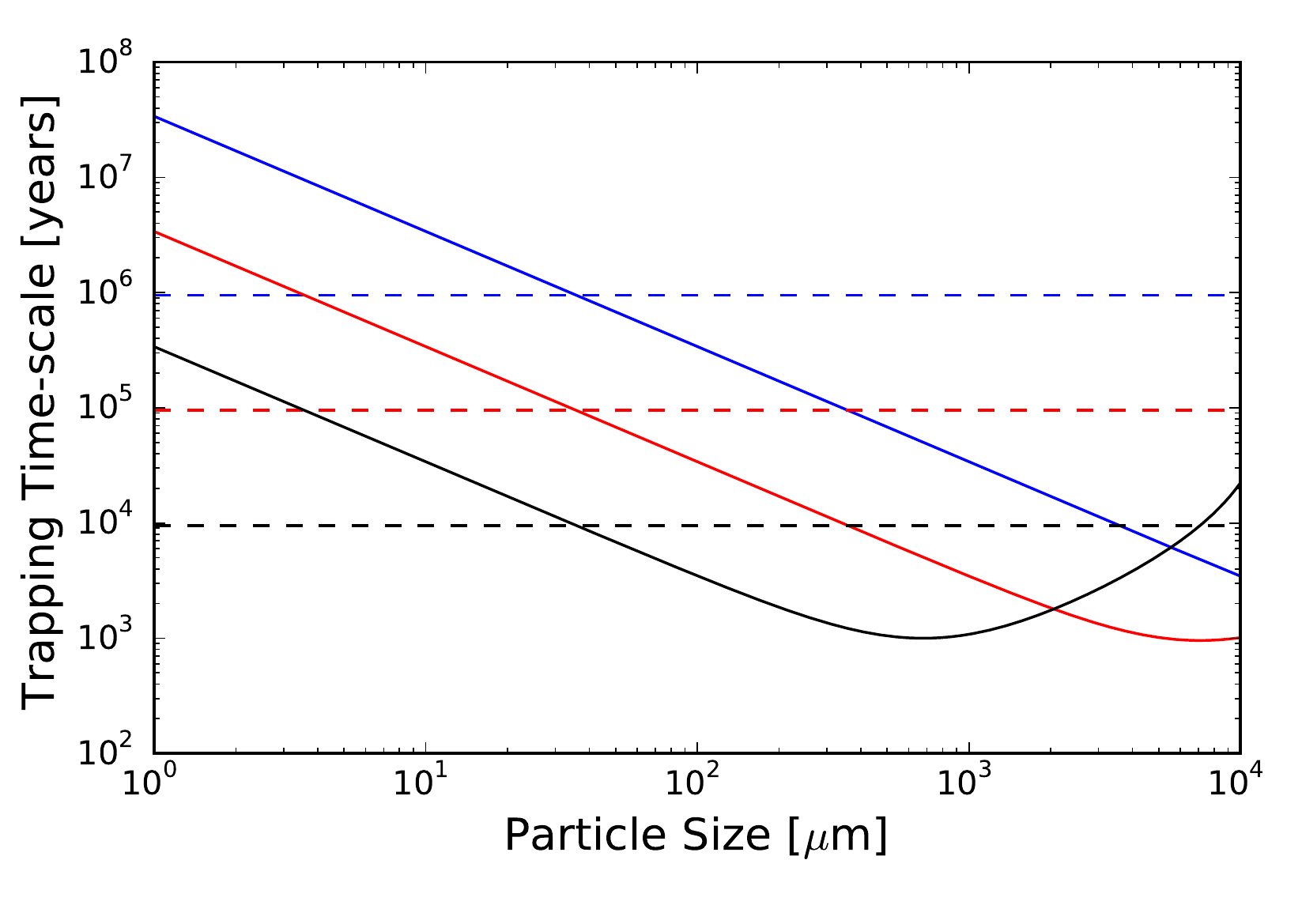}
\caption{The trapping time-scale as a function of particle size (solid) and gas advection time-scale (dashed) for $\alpha=10^{-4}$, 10$^{-3}$ \& 10$^{-2}$ (from top to bottom). Any particle that has a trapping time-scale shorter than the gas advection time-scale has a potential to become trapped in a pressure maxima on that time-scale. }\label{fig:trap_time}
\end{figure}

Assuming the dust particles quickly reach terminal velocity then their radial velocity is given by:
\begin{equation}
v_r=\frac{u_{\rm gas}/\tau_s +nv_K}{\tau_s+\tau_s^{-1}},
\end{equation}
where $u_{\rm gas}$ is the gas radial velocity and $\tau_s$ is the non-dimensional stopping time, taking a value $\tau_s=\pi\rho_d s/2\Sigma$ for small particles, where $\rho_d$ is the density of the dust particle and $s$ is the particle size. Taking typical transition disc numbers (i.e. $R=20~AU$, $\dot{M}=10^{-8}$~M$_{\odot}$~yr$^{-1}$, $H/R=0.1$, $M_*=1$~M$_\odot$) we can plot the trapping time-scale $|R/(v_r-u_{\rm gas})|$ as a function of particle size, the result of which is shown in Figure~\ref{fig:trap_time} for various choices of the viscous $\alpha$ parameter. 

This simple figure demonstrates large $\sim$ mm to cm sized grains have the potential to become trapped in the pressure maxima, and they will do so on a rapid time-scale of $10^3-10^4$~years. This is in good agreement with the mm continuum images that suggest that the dust particles are confined to a narrow ring, consistent with dust trapping \citep{Pinilla2012b} and furthermore, imaging a multiple wavelengths suggest that larger cm sized grains are more spatially confined compared to mm sized particles \citep{vanderMarel2015b,Casassus2015b}. Thus planet induced gaps can filter large sized dust particles \citep{pm2004,pm2006,Rice2006,Ayliffe2012,Zhu2012,Zhu2014,Pinilla2015} from the gas. Therefore, narrow rings in the mm appear to be a strong confirmation of the planet scenario.

\begin{figure}
\centering
\includegraphics[width=\columnwidth]{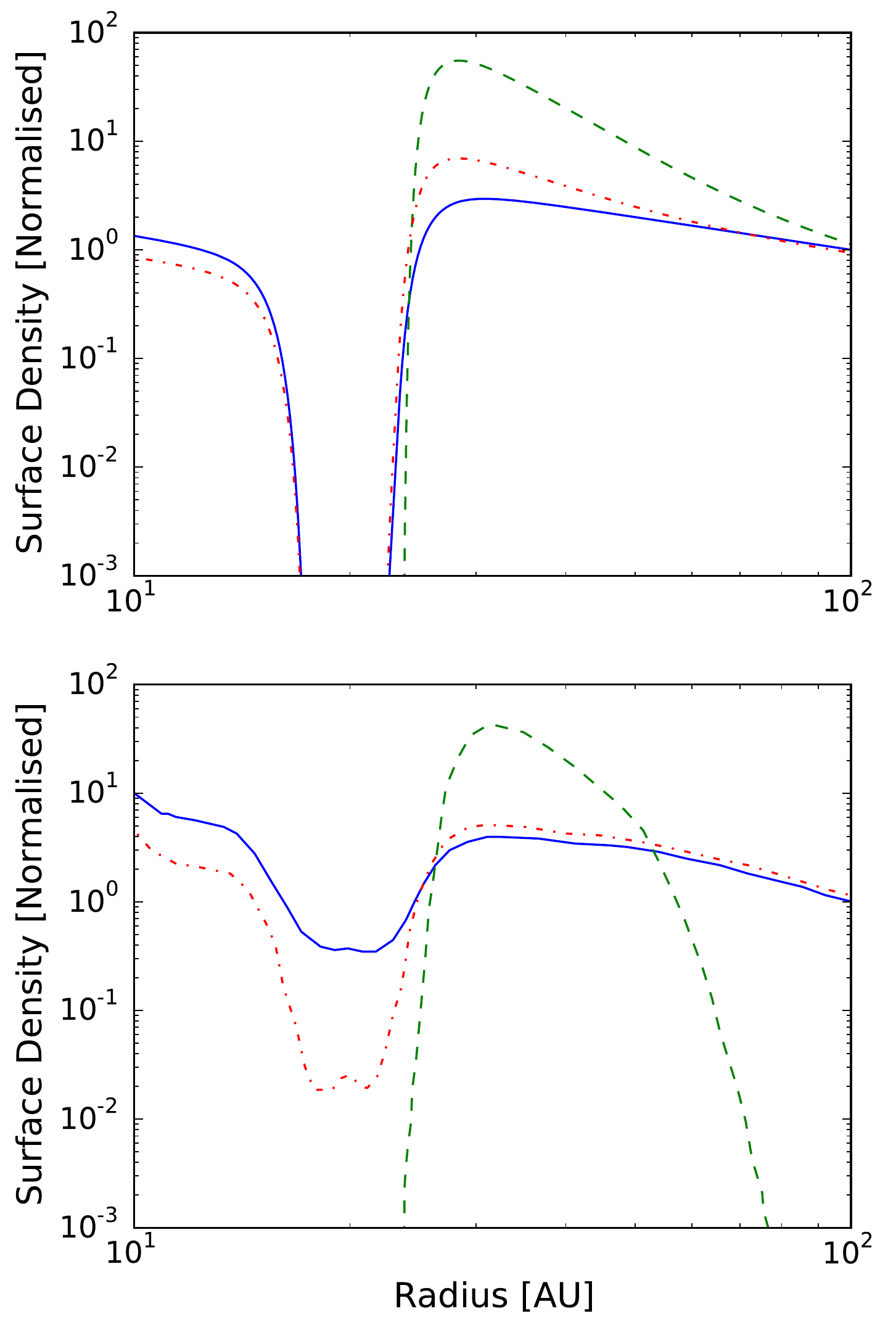}
\caption{The gas and dust surface density profiles for a 4~M$_J$ planet (top - taken from \citealt{Owen2014}) and a 3~M$_J$  planet 
(bottom - taken from \citealt{Zhu2012}) orbiting at 20~AU. The gas (solid),  1~mm size dust particles (dashed) and 0.03mm size dust particles (dot-dashed) are shown. \citet{Zhu2012,Owen2014} use different numerical schemes, assumptions about planet accretion, boundary conditions and viscosity profiles (for example \citet{Owen2014} has a constant flux of gas and dust at large radius and \citet{Zhu2012} does not)  yet conclusions about dust trapping remain robust: mm sized partciles are effectively trapped, small particles flow through the gap into the inner disc.}\label{fig:owen2014}
\end{figure}

However, Figure~\ref{fig:trap_time} also alludes to a problem for the planet trapping scenario. Small dust particles ($1-10$~$\mu$m) have trapping time-scales that are both longer than a disc's lifetime and many orders of magnitude longer than the gas advection time. Any particle that has a trapping time longer than the advection time-scale will be carried with the gas rather than moving towards the pressure maxima. Thus, dust filtration alone {\it cannot} remove small $1-10$~$\mu$m dust particles. This is problematic as small particles dominate the opacity, so if they indeed follow the gas through the gap and populate the inner disc then the disc will show a strong IR-excess at all wavelengths and have an SED consistent with a primordial disc. Such an expectation has been confirmed using numerical gas and dust simulations \citep{Zhu2012,Owen2014} who showed that while the mm rings could be explained easily, the small dust that percolated into the inner disc gave rise to a primordial SED. The surface density profile in gas and dust obtained by \citet{Zhu2012} (bottom) and \citet{Owen2014} (top) are shown in Figure~\ref{fig:owen2014}, where a few Jupiter mass planet at 20~AU embedded in a the gas disc. Clearly the sub-micron dust follows the gas across the gap giving rise to a primordial looking SED. 

The fact the small dust particles are not trapped has been argued as evidence for the planet scenario in the context of the recent SEEDS images that show scattered light from small dust particles inside the mm cavity. However, such an explanation does still not account for the lack of IR excess and ad-hoc small dust surface density profiles are required to explain all the observations. Furthermore, given the single case where MIR imaging has been performed (IRS48-\citealt{Geers2007}), which shows that the thermal emission at MIR wavelengths also shows a cavity this seems to indicate that the bulk of the small dust is also trapped along with the mm-sized dust particles. 

There have been a couple of attempts to reconcile the planet model with the lack of small dust particles in the inner disc. Several authors have suggested a chain of multiple planets could suppress the surface density over a large enough radial range to explain the lack of NIR emission, while still allowing the small dust particles to flow into the inner disc. \citet{Zhu2011} explored this using numerical simulations, finding that with 4 planets the surface density could be reduced enough to explain the SEDs of some pre-transition discs. However, in all simulations that were successful in explaining the lack of NIR excess the accretion rate was suppressed by a least two-orders of magnitude with respect to the observed value. Such a result is not unexpected, if every planet accretes a fraction $f$ of the incoming gas flow from outside its orbit then the total accretion rate from that outside the planet with the largest separation and the star is $f^N$, if it takes a minimum of 4 planets in order to suppress the surface density to necessary values and for an optimistically small $f=0.5$ (\citealt{Lubow2006} suggest $f\lesssim 1/3$) then the accretion rate onto the star is $\sim 5\%$ of that in the outer disc. Given the observed accretion rates can be  of order a few $10^{-8}$~M$_\odot$~yr$^{-1}$, this would suggest the accretion rate in the outer disc should be of order $10^{-7}$-$10^{-6}$~M$_\odot$~yr$^{-1}$. Typical primordial discs rarely show accretion rates this high, and sustaining them over a {\it minimum} required lifetime of a few $10^5$~years would require incredibly large disc masses (probably so large they would just destroy them-selves by fragmentation \citealt[e.g.][]{Lodato2005}). Therefore, a chain of multiple planets from 10s of AU to $\lesssim 1$~AU seems unlikely to represent the mm-bright transition discs.

\citet{Owen2014}, suggested that high measured mass accretion rates and implication that a significant fraction of the material entering the gap accreted onto the planet resulted in a large accretion luminosity of order:
\begin{eqnarray}
L_p&\approx&7\times10^{-3}{\rm L}_\odot ~f^{-1}\left(\frac{M_p}{3~{\rm M}_J}\right)\nonumber\\&\times&\left(\frac{\dot{M}_*}{10^{-8}~{\rm M~yr}^{-1}}\right)\left(\frac{R_p}{10^{10}~{\rm cm}}\right)^{-1},
\end{eqnarray}   
and will be the dominant UV luminosity source close to the gap edge. For small dust grains the opacity of an individual dust grains $\propto 1/a$, therefore radiation pressure will be important for small dust particles. \citet{Owen2014} demonstrated that for massive $M_p>3$~M$_J$ efficiently accreting $f\lesssim 1/2$ planets that radiation pressure could trap the small dust particles at the gap edge, while the accretion rate onto the star was large $\sim 10^{-8}$~M$_\odot$~yr$^{-1}$. Such simulations were performed in orbit averaged 1-D calculations and it remains to be seen if such a mechanism works in 3D. \citet{Owen2014} speculated that such planets would only appear as a transition disc with a strongly suppressed NIR excess at certain points during its evolution, i.e. at large radius when the accretion rate is high enough and the planet mass is large, whereas when the radiation pressure trapping was inefficient it would only trap the large dust particles by ordinary gas-drag controlled trapping, letting the small dust particles into the inner disc. \citet{Owen2014} did not perform any evolutionary calculations to quantify the evolutionary features of such a scenario. This model obviously makes a strong prediction that there should be a bright accreting planet source inside {\it all} mm-bright transition discs, which is at least true in one case {\bc \citep{Sallum2015}}. Additionally, there should be a large population of discs that have primordial SEDs and mm-cavities waiting to be discovered, that should have weaker mm-holes (smaller planet mass) and/or lower accretion rates (insufficient radiation pressure to trap dust). Currently, radiation pressure trapping by an accreting planet represents the only model that explains all the necessary observational characteristics of a mm-bright transition disc (accretion, mm cavity, no strong NIR excess). 

Finally, several authors \citet{Zhu2012,Pinilla2012b} have speculated in passing that grain-growth of the small dust particles that make it through the gap to large sizes (hence reducing the opacity) may provide the solution. In order for grain-growth to work, it requires tuning the various time-scales for dust evolution and drag. One requires that the growth times-cale for small particles is fast compared to their drift time-scale (such that the particles can grow to large sizes) at which point the drag time-scale for the large particles must be faster than the fragmentation times-cale. Otherwise the newly formed large particles will fragment and produce small particles again with a high opacity. Additionally, the mass in the particles that have grown to large sizes needs to be small that does not contain appreciable mm opacity, but high enough that the remaining small dust particle population also has a low opacity. Whether all these conditions can be met in the wide observed range of parameter space still remains an open question.     

\subsection{Evolution of a transition created by planet}

In contrast to building a model transition disc structure created by the planet, the evolution of a such a disc structure and how it relates to the {\it population} of transition discs has barely been studied. Specifically, there are limited predictions for how a planet induced transition disc should evolve in accretion rate, hole size and mm flux. {\bc At least two studies} have attempted to address this by dispensing with how the planet creates the specific dust structures in the protoplanetary disc which give rise to a transition disc structure, rather they just assumed, as commonly invoked that a single planet creates a transition disc like dust structure with a hole size similar to the separation of the planet. \citet{Clarke2013}, studied how the the planet-disc system would evolve when the planet migrates and \citet{Rosotti2015} studied how an evolving disc that contained a planet and was undergoing photoevaporation evolved. 

Unfortunately, both studies returned negative outlooks by finding that both models populated unobserved regions of the transition disc space and suggested that a simple model where a single planet that created a transition disc structure could not explain the observed population of (type b) transition discs. 

\subsubsection{Role of Migration}

\begin{figure}
\centering
\includegraphics[width=\columnwidth]{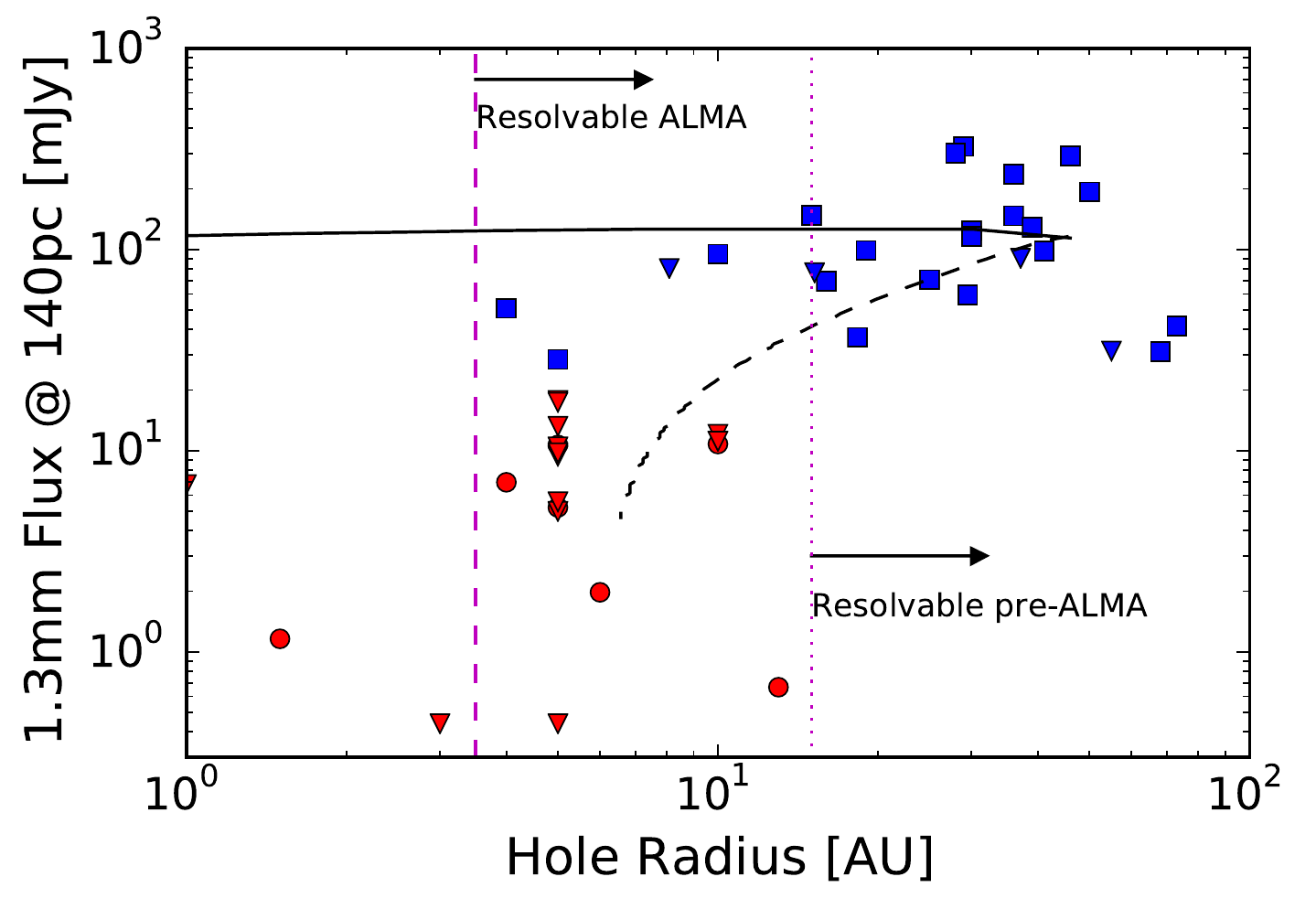}
\caption{The loci of a disc with a migrating planet in the inner hole radius and mm flux planet for a 10~M$_J$ (solid) and 100~M$_J$ companion perfomed by \citet{Clarke2013}. The points show observed transition discs taken from the sample described in Section~2 with mm-faint discs shown in red and mm-bright discs shown in blue; triangles show upper limits in mm flux. The vertical dotted line shows the hole radius of the smallest inner hole size directly imaged before ALMA. The vertical dashed line shows the hole radius resolvable with ALMA, assuming the same resolution as the HL-Tau observation \citep{HLTAU2015} at 140pc.}\label{fig:mm_fluxes}
\end{figure}

Firstly, \citet{Clarke2013} studied the evolution of an evolving disc with a migrating planet of various masses, as well as calculating observational diagnostics. Specifically, comparing the evolution of  discs in the mm-flux against hole size plane, one found that planetary mass objects migrate to small hole sizes by type II migration, but their mm fluxes did not decline appreciably. This gave rise to a population of transition discs with small $\lesssim 10$~AU transition discs and comparable mm fluxes to mm-bright transition discs. There are very few transition discs observed with large mm-fluxes and small hole sizes. The only way \citet{Clarke2013} could resolve this paradox was by making the ``planet'' sufficiently massive ($\sim 0.1$~M$_\odot$), such that its inertia prevented disc driven migration and stalled at large radius \citep{Syer1995}. We can quantify this result by considering that mm sized dust particles of mass $M_{\rm mm}$ are trapped in a ring at radius $R$ and width $\Delta R$, such that it has a surface density $\Sigma_{\rm mm}\sim M_{\rm mm}/R\Delta R$. The mm flux of this ring scales as:
\begin{equation}
F_{\rm mm}\approx R\Delta R B_{\rm mm} (T)\left[1-\exp\left(-\tau_{\rm mm}\right)\right],
\end{equation} 
where $B$ is the plank function and $\tau$ is the optical depth. Continuum mm dust emission is usually optically thin in protoplanetary discs and assuming the trapped dust mass does not fall as the planet migrates then the $F_{\rm mm}\propto B_{\rm mm}(T)\propto R_{\rm hole}^{-1/2}$ assuming the Rayleigh-Jean's limit. Therefore, the results of \citet{Clarke2013} are easy to understand, a migrating planet which traps all the mm sized dust outside its orbit does not fall in mm flux as it migrates in, unless it becomes very optically thick (even if it becomes optically thick it only falls weakly $\propto R_{\rm hole}^{1/2}$ for a $\Sigma\propto R^{-1}$ gas surface density profile). The observational data of 1.3 mm flux plotted against cavity radius is shown in Figure~\ref{fig:mm_fluxes}, with the track for a migrating 10~M$_J$ (solid) and 100~M$_J$ (dashed) companions shown.  Indicating that it migrates into the region of parameter space where very few transition discs are observed. The massive companions required to survive migration into the inner disc with high mm-fluxes (by expelling material to large radius, reducing the mm-flux) are well above the current sensitivity levels of direct imaging studies of transition discs which currently are readily capable of detecting $10$~M$_J$ objects \citep{Kraus2011,KI2012,Kraus2012}. Furthermore, if all mm-bright transition contained objects of 10s of Jupiter masses that migrated and ``parked'' at separations $\lesssim 5$~AU, this would certainly overpopulate the ``brown-dwarf desert'' which suggests the sub-stellar companion fraction at a separation of several AU is $\sim 1\%$\citep{Kraus2008}, an order of magnitude lower than the transition disc fraction. 

It is possible that whatever mechanism creates the lack of NIR emission for planet hosting transition discs at large radius fails once the planet migrates to small separations and \citet{Owen2014}'s radiation pressure trapping mechanism qualitatively would give rise to such a transition; however, it is unclear if it could quantitatively make the NIR emission ``re-appear'' in the correct range of parameters. Furthermore, it would mean there is a large population of discs which are mm-bright, show small cavities in the mm ($\lesssim 10$~AU) but have primordial looking SEDs. 

\subsubsection{Role of photoevaporation}

Photoevaporation could provide the truncation mechanism necessary to stop the planet migrating into the inner disc and several authors have explored photoevaporation as a mechanism to stop migration in the context of explaining the orbital separation distribution of massive exoplanets \citep{Alexander2012,Ercolano2015}. \citet{Rosotti2013,Rosotti2015} studied the impact of photoevaporation on a disc with an embedded planet. The result of including a planet was that photoevaporative clearing of the inner disc was triggered early (as the planet helps starve the inner disc while the planet was at large radius). \citet{Rosotti2013} argued that one could plausibly explain the accretion rates and hole sizes of some transition discs. \citet{Rosotti2015} then followed this up by considering the evolution of such a system and the type of transition disc population would be produced. 

Since photoevaporation was triggered earlier, then the mass remaining in the outer disc was larger than without the planet. In this case it took much longer for the outer disc to be dispersed. \citet{Rosotti2015} performed a population synthesis calculation and found that on average the disc spent a large (the majority) of its time as a non-accreting transition disc with a large hole. The problem was worst at low X-ray luminosities (i.e. low photoevaporation rates) as the fraction of disc material removed in the wind compared to viscous draining becomes smaller; therefore, one cannot simply invoke low photoevaporation rates to solve the problem (unless thermal sweeping was much more efficient than previously estimates, although this seems unlikely).

\citet{Rosotti2015} could not come up with a viable solution to this problem within the framework of the standard picture of a single planet interacting with a photoevaporating disc under the influence of standard type II migration. {\it As such the lack of observed {\bc optically thick} non-accreting transition discs appears to be a severe problem for the commonly invoked planetary hypothesis for transition discs. }  

\subsection{Summary}
\begin{figure*}
\centering
\includegraphics[width=0.98\textwidth]{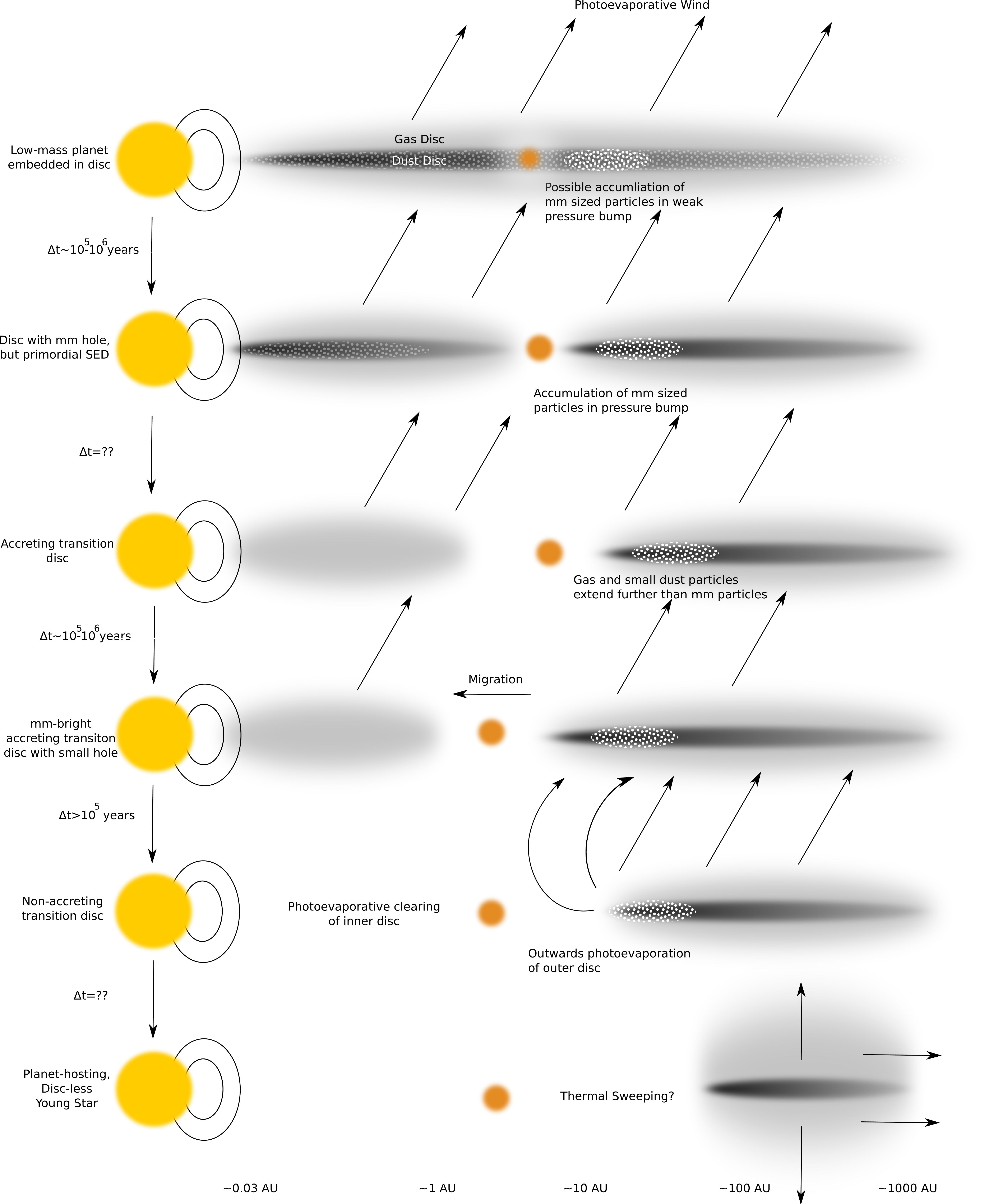}
\caption{Schematic picture of the evolution of a planet carved transition disc. Initially a low-mass planet embedded in the disc may trap mm-sized particles in a weak pressure trap (stage I). Once the planet is high enough mass to open a gap it will strongly trap mm-sized particles (stage II), dust still remains interior to the planet so it will appear as with primordial SED and mm cavity. Some (poorly understood) mechanism removes dust from the inner disc giving rise to an accreting transition disc (stage III). This planet will migrate and the disc will photoevaporate leading to a mm-bright transition disc with a small hole (stage IV) or after photoevaporation has opened a gap a non-accreting transistion disc (stage V). Eventually the remaining disc material is dispersed, possibily by thermal sweeping leaving a young planet-hosting, disc-less star (stage V). Stage IV and V are not observed.  }\label{fig:planet_diagram}
\end{figure*}

Giant planet formation takes place in protoplanetary discs and could under some circumstances give rise to a transition disc signature. Specifically, it is more well suited to the production of mm-bright transition discs: namely those with large mm fluxes, large holes and strong accretion. Furthermore, massive planets induce large pressure bumps in discs capable of trapping large amounts of mm particles and possibly provide a mechanism for creating the various types of non-axisymmetric structure seen recently \citep{Casassus2015c}. 

The evolutionary path of a giant planet forming disc that creates a transition disc is schematically shown in Figure~\ref{fig:planet_diagram}. A growing low-mass planet that cannot open a gap may trap particles in a weak pressure gradient \citep[e.g.][]{Dong2015b}. As the planet grows in mass it will eventually open a large gap, where the pressure gradient is strong enough to trap the mm-particles in a narrow ring on a short time-scale \citep[e.g.][]{Zhu2012,Pinilla2012b,Owen2014,Pinilla2015}. The small dust particles will persist in the inner disc shortly after gap opening, giving rise to a primordial SED but hole in the mm. Eventually, and through some process that is not fully understood (possibly grain growth \citealt{Zhu2012,Pinilla2012b} or radiation pressure trapping \citealt{Owen2014}) the inner disc becomes dust free giving rise to an accreting transition disc that could have a lifetime $\sim 10^{6}$~years. During this accreting transition disc phase the planet will migrate into the inner disc $\lesssim 10$~AU and the inner disc will be starved by photoevaporation. Migration gives rise to an accreting transition disc signature with a large mm flux and small hole, which is not observed \citep{Clarke2013}. The suppression of gas flow into the inner disc by the planet (through accretion and torques) triggers photoevaporation early \citep{Rosotti2013} giving rise to a non-accreting phase \citep{Rosotti2015} which is probably long lived and not observed (see Figure~\ref{fig:mdot_rhole}). Eventually the disc is completely dispersed, possibly by thermal sweeping.   

Like the photoevaporation model, when considered from an evolutionary standpoint it fails to completely describe the observed population of transition discs, predicting discs with large mm fluxes and small holes and non-accreting transition discs.

\section{OUTSTANDING ISSUES}
\label{sec:issues}
There several areas where the observational evidence is not clear and in some cases contradictory with our understanding of the disc structure. 
\subsection{Distribution of small dust particles}
The first {\bc outstanding issue concerns} the distribution and surface density of small dust particles inside the cavity. We know from the SEDs of transition discs - apart from small pile ups near the dust destruction front \citep{Espaillat2010} - that the cavities contain very few particles out to radii of $\sim 5-10$~AU. As well as the lack of IR excesses the strong 10~$\mu$m silicate features in many transition discs \citep{Kim2009,Espaillat2010,Kim2013} indicates that the cavity region is optically thin, and the inner edge of the dust hole is consistent with being irradiated directly by the star \citep{Ercolano2015}. However, the SEEDS images of several transition discs \citep[e.g.][]{Dong2012} indicate that the NIR scattered light (which must be coming from small particles high in the disc's atmosphere) show no evidence for a gap or a dip.  The fraction of the stars light intercepted (and then subsequently scattered) is - for grazing incidence -  proportional to the grazing angle ($\varphi$) given by \citep[e.g.][]{Chiang1997}:
\begin{equation}
\varphi=R\frac{\rm d}{\rm d R}\left(\frac{h_{\rm phot}}{R}\right),
\end{equation}
where $h_{\rm phot}$ is the height of the photosphere above the mid-plane. Assuming the small dust particles are well mixed with the hydrostatic gas disc, and the surface density of small dust particles were to rapidly change due to their removal at the edge of a planet gap, then the height of the photosphere would fall. This would cause a characteristic peak in the flaring angle as one crosses the gap edge. For the large surface density changes required to explain the lack of IR excess the change in grazing incident angle would almost certainly give rise to a bright ring in the scattered light emission, similar to that seen in the non-accreting transition disc J1604--2130 \citep{Mayama2012}. Therefore, at the moments their appears to be a contradiction between the scattered light images and SEDs. Part of the answer maybe that the peak of the mm image is not a good indicator of the gap edge and could be factor of $\sim 2$ larger \citep{Pinilla2012b} and due to their tighter coupling to the gas the small dust particles may be trapped close to the gap edge which is currently inside the SEEDS chronograph. 

\subsection{Gas in the cavity}

If we had detailed information on the distribution and kinematics of gas in the cavity the constrains on a given model would be much stronger. As mentioned previously, the fact the most transition discs are accreting indicates that the material accreting onto the central star must be resupplied. There have been several recent studies, mainly using {\it ALMA} that have aimed to quantify the properties of the gas disc, with several results indicating high dust-to-gas ratios in the outer disc suggestive of dust trapping \citep{Bruderer2014}. However, it is observations of gas in the cavity that are the most intriguing.  \citep{Bruderer2014,vanderMarel2015} used {\it ALMA} CO observations of several transition discs combined with photo-chemical modelling to determine the gas surface density profile of the transition disc and \citet{Carmona2014} used  CO $\rho$-vibrational lines to derive a gas surface density profile for the disc. In almost all cases these gas profiles showed that the outer gas disc extended inside the peak of the mm emission as expected from gas pressure driven particle trapping \citep{Pinilla2012b,Zhu2012,Zhu2014,Owen2014}. However, the  gas observations further indicated significant depletion of the gas in the inner disc, in some cases by many orders of magnitude. 
Since the majority of these discs are accreting onto the central star and the inner gas discs are likely to be the source of the accreting material we can calculate the draining time of these inner gas discs $t_{\rm drain}$. We do that here by integrating the inferred gas surface density profile upto the drop in gas surface density, to obtain the gas mass contained in the inner gas disc. Dividing the inner gas mass by the accretion rate gives the draining time for these inner discs. For some transition discs this exercise yields incredibly short draining time-scales: for DoAr44 \citep{vanderMarel2015c} one finds a timescale of $\sim 30$~years, for IRS48 \citep{Bruderer2014,vanderMarel2015c} a time-scale of $\sim 100$~years, for SAO20645 \citep{Carmona2014,vanderMarel2015c} a time-scale of $\sim 1000$~years and for RXJ1615-3255 \citep{vanderMarel2015} a time-scale $\lesssim 10^{4}$~years  and  for HD142157 \citep{Perez2015} a time-scale of $2\times10^{4}$~years, although some others do show more sensible time-scales e.g. LkCa 15 which has an inner disc draining time-scale of $\sim 10^6$~years. Clearly all most all these time-scales are incompatible with viscous draining which would point to a time-scale more like $10^6$~years for hole radii of several 10s of AU. 

Infact HD142157 shows evidence for fast radial inflows \citep{Casassus2013,Rosenfeld2014,Casassus2015a}, which would lead to the rapid replenishment time-scales. However, one must caution that HD142157 maybe a very unique object: it has an extremely large hole size of 140~AU \citep{Perez2015} and is believed to possess a large inclination or warp with respect to a more compact inner disc with an inclination change of 70 degrees \citep{Marino2015}. Such a dramatic inclination change is certainly not present in the other discs with short replenishment time-scales. Since there are no observed non-accreting transiting discs we are not simply catching these discs just before the inner disc completely drains onto the central star. Therefore, these inner gas discs must be replenished from the massive gas discs outside the cavity (which certainly possess enough mass to supply the observed accretion over Myr time-scales, based on the inferred disc masses). However, a mechanism that can refill these inner disc cavities at such rapid time-scales.  In the case of IRS 48 this resupply time-scale is worrying similar to the orbital time-scale at the cavity radius as it would require highly super-sonic replenishment.

\section{INSIGHTS FOR PROGRESS}
\label{sec:progress}
While much progress has been made in the decade since transitions discs became an important area of study, it appears we are still a long way off fully understanding the nature of transition discs. It is know clear that there are probably at least two distinct populations of transition discs: one at with low mm-fluxes (type a), small hole sizes and low accretion rates and one with high mm-fluxes (type b), large hole sizes and high accretion rates. It is tempting to assign each of these different types to the two popular mechanisms for creating transition discs: i.e. mm-faint transition discs are created by disc undergoing dispersal by photoevaporation and mm-bright are transition discs that are undergoing interactions with embedded planets. 

While the models can successfully reproduce the observed characteristics of single objects when considered in isolation. When considered as a population the models tend to produce too many non-accreting transition discs with large holes, of which there are only a few known examples: SR21 and J1604-2130 both of which have large holes $\gtrsim 20$~AU \citep{Andrews2011,Zhang2014} and no detectable accretion rate \citep{Manara2014}. While in comparison there are many $\gtrsim 20$ transition discs with large holes that have accretion rates $\sim 10^{-9}-10^{-8}$~M$_\odot$~yr$^{-1}$.

\subsection{Improvements in the theoretical models}

The photoevaporation model has {\bc predominately} been studied from an evolutionary standpoint. \citet{Alexander2007} demonstrated that once the gap opened dust drag quickly cleared the inner disc giving rise to an accreting transition disc phase and \citet{Owen2011,OC12} showed that the X-ray photoevaporation model was broadly consistent with the range of accretion rates and hole sizes shown by mm-faint transition discs. However, as the majority of the remaining disc mass after gap opening is present at large radius, the photoevaporation model still predicts that their should be a large fraction of non-accreting transition discs with large hole sizes. Thermal sweeping \citep{Owen2013} may go some way to alleviate this problem by dispersing the outer disc through a rapid, dynamic instability. However, the mechanism is yet to be fully explored and a detailed analysis of the photoevaporation and thermal sweeping model compared to the full distribution of observed transition disc is yet to be performed.

Comparably there have been relatively very few studies of the evolution of a planet driven transition disc. Those studies have thrown up problems with the idea. The ``planetary-mobility problem'' discussed by \citet{Clarke2013} indicated that planet would migrate into a forbidden region of planet space, giving rise to a transition disc with a high mm flux and small hole. Companions that were too massive to migrate (\citealt[e.g.][]{Syer1995}) into the inner disc required masses $\gtrsim 0.1$~M$_\odot$, which are well above the detection limits from direct imaging \citep{Kraus2011,KI2012,Kraus2012}. The ``planetary-mobility problem'' could easily be solved by postulating that the process that gives rise to a transition disc SED fails in the inner disc; \citet{Owen2014} postulated radiation pressure from an accreting proto-planet could work, although whether this works has yet to be full assessed. Furthermore, \citet{Rosotti2015} showed that when one includes disc dispersal by photoevaporation one always produces a set of long lived non-accreting transition discs with large holes, and is difficult to find a solution to this problem. \citet{Rosotti2015} argued that even if giant planets don't create a transition disc signature while the star is accreting, once photoevaporation clears the inner disc, the disc will appear as a long lived non-accreting transition disc.  

There are still issues how the planet sculpts the dust distribution to give rise to a transition disc signature at all wavelengths, not just the NIR or mm separately. One must find a model that produces a mm ring, scattered light image consistent with the SEEDs structure, a lack of NIR excess and active accretion. Several models are beginning to tackle these issues, however, there is no clear paradigm that works. Perhaps most difficult problem to explain is how a planet removes enough small dust such that it explains the lack of NIR excess in the inner disc. Several authors have hypothesised that the trapping of mm-sized particles might promote the growth of the small particles to large ones. Calculations by \citet{Pinilla2012b,Pinilla2015} have demonstrated that this could in principle work; however, since these particle may drift, grow and fragment on a time-scale comparable to the lifetimes of transition discs then how this removal mechanism causes the NIR colours and scattered light images to evolve in time is of crucial importance to asses whether grain growth can provide the answer. 

Finally, if one is truly postulating that the majority of transition discs contain planets, one cannot ignore the constraints from exoplanet studies. If we assume that all mm-bright transition discs are long lived and have lifetimes comparable to the total disc lifetime (i.e. they look like transition discs for their entire life-time - minimising the fraction of planet hosting transition discs) then the giant planet hosting, transition disc fraction is $\sim 10\%$.     \citet{Cumming2008} estimated the giant planet fraction to be $\sim 10\%$; finding it to be a decreasing function of planet mass \citep{Cumming2008,Brandt2014} with the majority of planets to be dominated by lower $\lesssim$~Jupiter mass objects. Since, many of the models for transition discs invoke massive  planets ($\sim 5- 10$~M$_J$) \citep{Pinilla2012b,Pinilla2015,Dong2015}, there appears to be slight tension with the exoplanet statistics as the planet fraction at these masses maybe closer to $\sim 1-3\%$ \citep{Brandt2014}. However, without proper evolutionary calculations one cannot directly compare. The exoplanet statistics will only improve with time and theoretical models that invoke planets should begin to consider the constraints the exoplanet population places on their models. 

\subsection{Areas of more detailed observational studies}
While observations of transition discs have grown rapidly over the last decade, they have often been highly biased. For example almost all the imaging has concentrated on mm-bright objects and have thus necessary targeted mm-bright transition discs. Now we have entered the {\it ALMA} and higher spatial resolution and deeper studies are possible mm-faint transition discs should be an area of improved study. For example do mm-faint transition discs show narrow mm-rings or more discs outside the cavity, or do they even show a cavity at all? The photoevaporation creates a much less sharp gap than a planet, as such particle trapping should be less efficient and one might expect more extended mm emission from outside any cavity. While mm-faint transition discs appear to have the lowest mm flux all protoplanetary discs, do they also show the same depletion in gas as would be expected from a disc dispersal model that creates transition discs. 

Furthermore, more information on the gas content in the cavity will be crucial to identifying the mechanism that creates transition disc structures. Current observations of several transition discs suggest surprisingly short inner gas disc lifetimes based on the observed accretion rates that are certainly not consistent with a viscous process as discussed in section~\ref{sec:planet}. Quantifying whether these low inner gas disc masses are consistent with rapid radial transport \citep[e.g.][]{Rosenfeld2014,Casassus2015a} or some other kinematic signatures will place strong constraints on the transition disc mechanisms. Also quantifying the amount of small dust inside the cavity, both through scattered light and thermal (MIR) imaging would help unravel the mechanism that not only traps the mm-sized particles but also removes the small particles in the inner disc is crucial. 

Finally, it must be emphasised that we cannot understand transition discs if we do not have an understanding of the general properties of primordial discs with which to compare. For example, are the non-axis symmetric structure we see in transition discs purely associated with this type of object or just a feature of all protoplanetary discs? There are several objects that show mm-cavities but primordial discs (MWC 758, WSB 60, \citealt{Andrews2011}). Whether these are an isolated rare example or point to a large hidden population of discs with no NIR deficient and a mm hole is an intriguing question. {\bc It cannot be emphasised enough that an unbiased {\it ALMA} survey of the disc population in nearby star forming regions is needed answer this question, as well a placing well known transition discs into context. Future insights into the evolution transition of discs can only placed into context if the underlying population of all discs is understood.} Characterising whether these disc are an a stage in the evolution of a transition disc: i.e. are they precursors or former transition discs would place further constraints on any transition disc models. For example these could correspond to discs hosting planets that are able to trap mm sized particles in a pressure trap, but unable to remove the small grains from the inner disc. Current observational facilities and instrumentation are capable of answering these questions in the near future and we may finally reach a point where we can use the observations of transition disc to {\it observationally} constrain planet formation.

\section{CONCLUSIONS}
\label{sec:conclusions}

Transition discs represent a rare class ($\sim 10\%$) of protoplanetary discs that show a significant deficit of NIR opacity when compared to standard primordial discs, but a MIR excess consistent with an optically thick disc. The interpretation that these discs  possess large holes (from 0.1-$>$ 1~AU)  has largely been born out by sub-mm imaging campaigns of the largest holes ($>15$~ AU) which shown mm rings outside the cavity. However, this does not mean this hole is devoid of gas, and the majority of the observed transition discs are accreting at rates comparable, but on average slightly lower than nominal T-Tauri stars, indicating the gas accreting onto the star must be resupplied from some reservoir. Very few $\lesssim 10\%$ of the observed transition discs are non-accreting transition discs with large holes $\gtrsim 20$~AU and this fact is one of the strongest constraints on any transition disc model. 

Additionally, the transition discs can be separated into two distinct groups: mm-faint transition discs have very low-mm fluxes (the lowest of all discs), have small hole sizes $\lesssim 10$~AU, low accretion rate $\lesssim 10^{-9}$~M$_\odot$~yr$^{-1}$ and they appear randomly drawn in spectral type from all discs. Such characteristics are consistent with what one would expect from young stars that are currently undergoing disc dispersal, transitioning between a disc-bearing and disc-less state. MM-bright transition discs have very high-mm fluxes (often the highest of all protoplanetary discs), have large hole sizes $\gtrsim 20$~AU, high accretion rates  $\sim 10^{-8}$~M$_\odot$~yr$^{-1}$ and are biased towards occurring around earlier-type stars. As such mm-bright transition discs appear inconsistent with the concept a young star that is caught in the act of dispersing its disc. 

We show that while mm-faint transition discs are consistent with the photoevaporation model and mm-bright transition discs could be consistent with embedded planets on an individual level, when compared from an evolutionary point of view to the {\it population} of transition discs problems arise. Specifically photoevaporation over predicts the number of {\bc optically thick} non-accreting transition discs and embedded planets over predicts both the number of accreting transition discs with small holes and high-mm fluxes and the number of non-accreting transition discs.  ``Thermal Sweeping'' may provide the answer to the lack of non-accreting transition discs; however, as yet it remains unproven. {\bc Furthermore, the link between accreting optically thick transition discs and optically thin IR excesses around WTTs remains to be investigated from a modelling perspective.} 

Additionally, some of the observations appear contradictory at the moment. In several cases there is no evidence for the depletion of small particles inside the cavity through scattered light images, yet the discs must possess a tiny amount of small dust in the inner cavity to match the SED and the strong 10$\mu$m silicate feature.  A large fraction of the observed inner gas discs  appear to have very rapid draining time-scales based on their current accretion rates, these time-scales are inconsistent with viscous diffusion.

There is still much observational and theoretical work that needs to be done, before we finally understand the origin and {\it evolution} of transition discs and can use them as an observational probe of planet formation. The order of magnitude increase in resolution and sensitivity for gas and dust imaging provided by {\it ALMA} may provide the key, provided we also attempt to understand primordial discs as well.

\begin{acknowledgements}
We thank the referee for comments that improved this review.
JEO acknowledges support by NASA through Hubble
Fellowship grant HST-HF2-51346.001-A awarded by the Space Telescope Science Institute, which is operated by the Association of Universities for Research in Astronomy, Inc., for NASA, under contract NAS 5-26555. JEO is grateful to F. Adams, R. Alexander, S. Andrews, P. Armitage, T.  Birnstiel, S. Casassus, E. Chiang, C. Clarke, R. Dong, C. Dullemond, B. Ercolano, C. Espaillat, C. Koepferl, M. Lin, G. Lodato, W. Lyra, C. Manara, N. van der Marel, S. Mohanty, I. Pascucci, P. Pinilla, D. Price, G. Rosotti, Y. Wu, Z. Zhu and participants of the Facebook ``Circumstellar Disks and Planet Formation'' group for fascinating discussions.      
\end{acknowledgements}

% UNCOMMENT THE LINES BELOW IF YOU WISH TO USE BIBTEX
\bibliographystyle{apj}
%\bibliography{yourbibfile}

\end{document}